\documentclass[10pt,twocolumn,letterpaper]{article}

\usepackage{iccv}
\usepackage{times}
\usepackage{epsfig}
\usepackage{graphicx}
\usepackage{amsmath}
\usepackage{amssymb}
\usepackage{booktabs}

\usepackage{epsfig}
\usepackage{ dsfont }
\usepackage{ cite }
\usepackage{float}
\usepackage{color}
\usepackage{xcolor}
\usepackage{soul}
\usepackage{float}
\usepackage{url}
\usepackage{afterpage}
\usepackage{rotate}
\usepackage{makecell}
\usepackage{multirow}
\usepackage{tabularx}
\usepackage{algorithm}
\usepackage{algorithmicx}
\usepackage{algpseudocode}
\usepackage{mathtools}
\usepackage{etoolbox}
\usepackage{lipsum}
\usepackage{bbm}
\usepackage{listings} 
\usepackage{tabularx}
\usepackage{subfig}

\usepackage[accsupp]{axessibility}

\algdef{SE}[SUBALG]{Indent}{EndIndent}{}{\algorithmicend\ }%
\algtext*{Indent}
\algtext*{EndIndent}

\newcommand{\redb}[1]{\textcolor{red}{\textbf{#1}}}
\newcommand{\blueb}[1]{\textcolor{blue}{\textbf{#1}}}

\newcolumntype{Y}{>{\centering\arraybackslash}X}
\algnewcommand{\nComment}[1]{\Statex \Comment{#1}}

\usepackage[pagebackref=true,breaklinks=true,letterpaper=true,colorlinks,bookmarks=false]{hyperref}
\usepackage[capitalize]{cleveref}
\crefname{section}{Sec.}{Secs.}
\Crefname{section}{Section}{Sections}
\Crefname{table}{Table}{Tables}
\crefname{table}{Tab.}{Tabs.}

\iccvfinalcopy 

\def\iccvPaperID{4129} 

\ificcvfinal\fi

\begin{document}

\title{BoMD: Bag of Multi-label Descriptors for Noisy Chest X-ray Classification}

\author{
\parbox{0.7\linewidth}{\centering
Yuanhong Chen \textsuperscript{\rm 1} $\thanks{First two authors contributed equally to this work.}$ $\quad$
Fengbei Liu \textsuperscript{\rm 1} \footnotemark[1] $\quad$
Hu Wang \textsuperscript{\rm 1} $\quad$
Chong Wang \textsuperscript{\rm 1} $\quad$
Yuyuan Liu \textsuperscript{\rm 1} $\quad$
Yu Tian\textsuperscript{\rm 2} $\quad$
Gustavo Carneiro\textsuperscript{\rm 3} $\newline$ 
\textsuperscript{\rm 1} Australian Institute for Machine Learning, University of Adelaide \\
\textsuperscript{\rm 2} Harvard Medical School, Harvard University \\
\textsuperscript{\rm 3} Centre for Vision, Speech and Signal Processing, University of Surrey}
}

\maketitle
\ificcvfinal\thispagestyle{empty}\fi

\begin{abstract}
Deep learning methods have shown outstanding classification accuracy in medical imaging problems, which is largely attributed to the availability of large-scale datasets manually annotated with clean labels. 
However, given the high cost of such manual annotation, new medical imaging classification problems may need to rely on machine-generated noisy labels extracted from radiology reports. 
Indeed, many Chest X-Ray (CXR) classifiers have been modelled from datasets with 
noisy labels, but their training procedure is in general not robust to noisy-label samples, leading to sub-optimal models.
Furthermore, CXR datasets are mostly multi-label, so current multi-class noisy-label learning methods cannot be easily adapted.
In this paper, we propose a new method designed for noisy multi-label CXR learning, which detects and smoothly re-labels noisy samples from the dataset to be used in the training of common multi-label classifiers. 
The proposed method optimises a bag of multi-label descriptors (BoMD) to promote their similarity with the semantic descriptors produced by language models from multi-label image annotations.
Our experiments on noisy multi-label training sets and clean testing sets show that our model has state-of-the-art accuracy and robustness in many CXR multi-label classification benchmarks, including a new benchmark that we propose to systematically assess noisy multi-label methods. Code is available at \url{https://github.com/cyh-0/BoMD}.
\vspace{-10pt}
\end{abstract}

\begin{figure}
    \centering
    \vspace{-5pt}
    \includegraphics[width=1.0\linewidth]{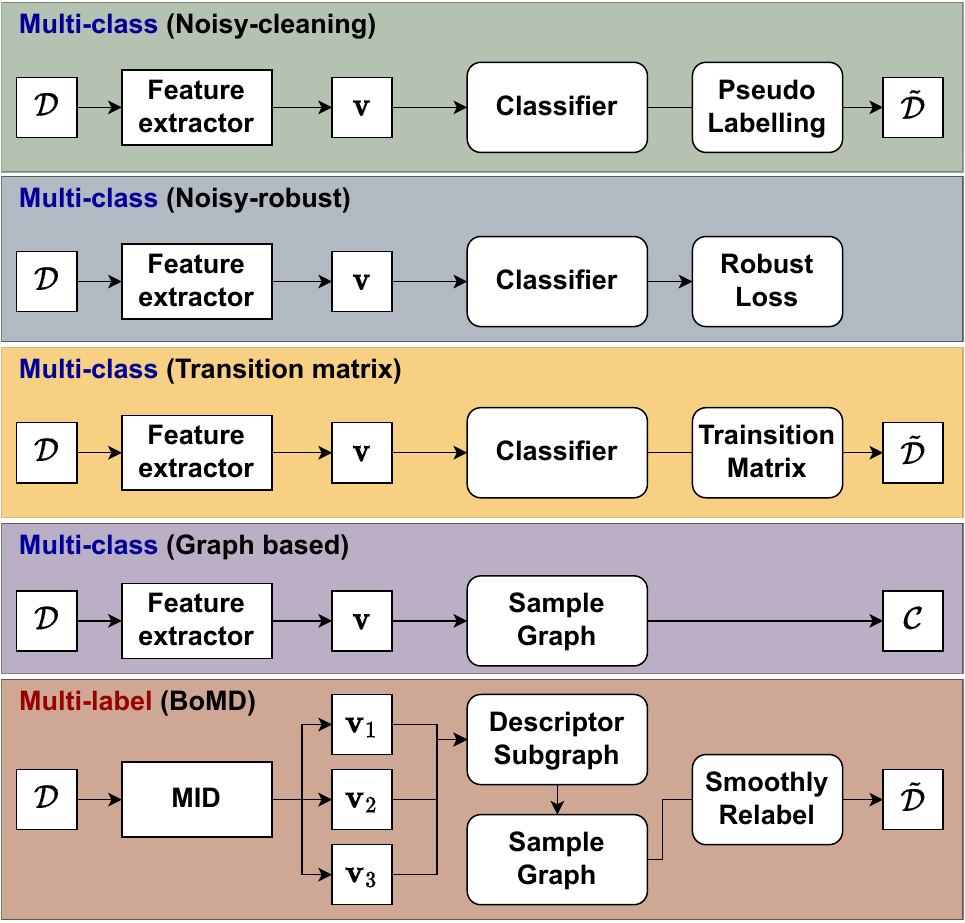}
    \caption{
    Comparison of multi-class LNL methods~\cite{li2020dividemix,liu2020early,bae2022noisy,iscen2022learning} and our noisy multi-label approach, BoMD, where the feature extractor returns a single descriptor $\mathbf{v}$ per image,
    $\mathcal{D}$ is the noisy training set, $\mathcal{C}$ is the clean set, and $\mathcal{\tilde{D}}$ is the re-labelled training set.
    BoMD has two components: 1) learning of a bag of multi-label image descriptors (MID) $\{\mathbf{v}_1,\mathbf{v}_2,\mathbf{v}_3\}$ to represent the image, and 2) smooth re-labelling of images driven by a graph structure based on the fine-grained relationships between MID descriptors.
    }
    \label{fig:motivation}
    \vspace{-10pt}
\end{figure}

\section{Introduction}

The promising results produced by deep neural networks (DNN) in medical image analysis (MIA) problems~\cite{litjens2017survey} is attributed to the availability of large-scale datasets with accurately annotated images.
Given the high cost of acquiring such datasets, the field is considering more affordable automatic annotation processes by Natural Language Processing (NLP) approaches that extract multiple labels (each label representing a disease) from radiology reports~\cite{nih,cxp}. 
However, mistakes made by NLP combined with uncertain radiological findings can introduce label noise~\cite{oakden2017exploring,oakden2020exploring}, as can be found in NLP-annotated Chest X-ray (CXR) datasets~\cite{nih,cxp} whose noisy multi-labels can mislead supervised training processes.
Nevertheless, even without addressing this noisy multi-label problem, current CXR multi-label classifiers~\cite{rajpurkar2017chexnet,ma2019multi,hermoza2020region,ben2021semantic} show promising results. 
Although these methods show encouraging multi-label classification accuracy, there is still potential for improvement that can be realised by properly addressing the noisy multi-label learning problem present in CXR datasets~\cite{nih,cxp}.

Current learning with noisy label (LNL) approaches focus on leveraging the hidden clean-label information to assist the training of DNNs (see Fig.~\ref{fig:motivation}). 
This can be achieved with techniques that clean the label noise~\cite{li2020dividemix, kim2021fine}, robustify loss functions~\cite{liu2020early,liu2021nvum,iscen2022learning}, estimate label transition matrices~\cite{goldberger2016training,yao2020dual,xia2020part,bae2022noisy}, smooth training labels~\cite{lukasik2020does, zhang2021delving, wei2021smooth}, and use graphs to explore the latent structure of data.~\cite{wu2020topological,wu2021ngc,iscen2022learning}. 
These methods have been designed for noisy multi-class problems and do not easily extend to \textit{noisy multi-label} learning, which is challenging given the potential multiple label mistakes for each training sample. 
In addition, a key characteristic of multi-label classification is the inherent positive-negative imbalance~\cite{ridnik2021asymmetric} issue. Such an issue may cause sample-selection based methods (e.g., DivideMix~\cite{li2020dividemix}) to select an extremely imbalanced clean set, where the majority of identified clean samples belong to the 'No Findings' class. Additionally, it could impede the accurate estimation of the posterior probabilities for the noisy or intermediate classes~\cite{li2022estimating}."
To the best of our knowledge, the state-of-the-art (SOTA) approach that handles noisy multi-label learning is NVUM~\cite{liu2021nvum}, which is based on an extension of early learning regularisation (ELR)~\cite{liu2020early}.
NVUM shows promising results, but it is challenged by the different early convergence patterns of multiple labels, which can lead to poor performance for particular label noise conditions, as shown in our experiments.
Additionally, NVUM is only evaluated on real-world CXR datasets~\cite{nih,cxp} without any systematic assessment of robustness to varying label noise conditions, preventing a more complete understanding of its functionality.

In this paper, we propose a new solution specifically designed for the noisy multi-label problem by answering the following question: \textbf{\textit{can the detection and correction of noisy multi-labelled samples be facilitated by leveraging the semantic information of training labels?}} available from language models~\cite{peng2019transfer,huang2019clinicalbert,lee2020biobert}?
This question is motivated by the successful exploration of language models in computer vision~\cite{couairon2022embedding,hu2018learning,zhang2016fast,ben2021semantic}, with methods that leverage semantic information to influence the training of visual descriptors; an idea that has not been explored in noisy multi-label classification. To answer this question, we introduce the 2-stage Bag of Multi-label Descriptors (BoMD) method  (see~\cref{fig:motivation}) to smoothly re-label noisy multi-label image datasets that can then be used for training common multi-label classifiers.
The first stage trains a feature extractor to produce a bag of multi-label image descriptors by promoting their similarity with the semantic embeddings from language models. 
For the second stage, we introduce a novel graph structure, where each image is represented by a sub-graph built from the multi-label image descriptors, learned in the first stage, to smoothly re-label the noisy multi-label images.
Compared with graphs built directly from a single descriptor per image~\cite{iscen2022learning}, our graph structure with the multi-label image descriptors has the potential to capture more fine-grained image relationships, which is crucial to deal with multi-label annotation.
We also propose a new benchmark to systematically assess noisy multi-label methods. In summary, our \textbf{contributions} are:
\begin{enumerate}
    \item A novel 2-stage learning method to smoothly re-label noisy multi-label datasets of CXR images that can then be used for training a common multi-label classifier;
    \item  A new bag of multi-label image descriptors learning method that leverages the semantic information available from language models to represent multi-label images and to detect noisy samples; 
    \item A new graph structure to smoothly re-label noisy multi-label images, with each image being represented by a sub-graph of the learned multi-label image descriptors that can capture fine-grained image relationships;
    \item The first systematic evaluation of noisy multi-label methods that combine the PadChest~\cite{padchest} and Chest X-ray 14~\cite{nih} datasets.
\end{enumerate}
We show the effectiveness of our BoMD on a benchmark that consists of training with two noisy multi-label CXR datasets and testing on three clean multi-label CXR datasets~\cite{liu2021nvum}. 
Results show that our approach has more accurate classifications than previous multi-label classifiers developed for CXR datasets and noisy-label classifiers.
Results on our proposed benchmark show that BoMD is generally more accurate and robust than competing methods under our systematic evaluation.

\section{Related Works}

\subsection{CXR multi-label classification}

Recently, we have seen many CXR multi-label classifiers being proposed, such as 
the CXR pneumonia detector~\cite{rajpurkar2017chexnet}.
Ma et al.~\cite{ma2019multi} introduce a new cross-attention network to extract meaningful representations. Hermoza et al.~\cite{hermoza2020region} propose a weakly-supervised method to diagnose and localise diseases.
Although these methods show promising results, there is still potential for improvement that can be realised by addressing the noisy multi-label learning of CXR datasets~\cite{nih,cxp}.

\subsection{Learning with Noisy Labels}
\noindent \textbf{Noise-cleaning methods} focus on detecting noisy samples.
For instance, Han et al.~\cite{han2018co} rely on the small-loss trick (i.e., clean samples have small losses) to co-teach two models.
Huang et al.~\cite{huang2019o2u} detect noisy samples that have unstable prediction by switching learning rates. 
Bahri et al.~\cite{bahri2020deep} discard samples whose labels disagree with a KNN classifier prediction.
Noise-cleaning methods can be combined with semi-supervised learning~\cite{berthelot2019mixmatch} to perform both the detection and correction of corrupted data. For example, DivideMix~\cite{li2020dividemix} removes the labels of samples classified as noisy and runs a semi-supervised learning method~\cite{berthelot2019mixmatch,liu2022translation}. 
FINE~\cite{kim2021fine} proposes a robust method to detect noisy samples by verifying the alignment of image features and class-representative eigenvectors.
Noise-cleaning methods generally employ two divergent networks to reduce confirmation bias~\cite{li2020dividemix,liu2022perturbed}, which substantially increases computational complexity.
Additionally, it is unclear if these methods can handle noisy multi-label problems since they do not in general capture fine-grained image relationships.

\noindent \textbf{Noise-robust methods} rely on robust loss functions to balance the overfitting effects caused by label noise in the training process. 
Early papers, such as~\cite{wang2019symmetric}, explore the symmetric property of cross-entropy (CE) loss for noise-robust learning. 
Zhang et al.~\cite{zhang2018generalized} propose the combination of Mean-Absolute-Error (MAE) and cross-entropy (CE) loss to achieve a good balance between convergence and generalisation. 
Ma et al.~\cite{ma2020normalized} show that 
any loss function can be robust to label noise by applying a simple normalization term. Recently,~\cite{englesson2021generalized} proposes a noise-robust Jensen-Shannon divergence (JSD) loss based on a soft transition between MAE and CE losses. Even though these methods can reduce overfitting effects, they also tend to under-fit the training data. This issue has been partially addressed by the early learning regularisation (ELR)~\cite{liu2020early} that proposes a regularization term which restricts the gradient from samples with corrupt labels. The non-volatile unbiased memory (NVUM) \cite{liu2021nvum} extents ELR to noisy multi-label problems. Although promising, ELR and NVUM are challenged by the different early convergence patterns of
multiple labels, which can lead to poor performance for particular label noise conditions, as shown in the experiments.

\noindent \textbf{Transition matrix methods} estimate the transition probability between clean and noisy labels. Goldberger et al.~\cite{goldberger2016training} propose a noise adaptation layer to estimate label transition. 
Yao et al.~\cite{yao2020dual} estimate the transition matrix using an intermediate class and a factorised matrix. Xia et al.~\cite{xia2020part} estimate part-dependent transition matrix for complex noise conditions. Bae et al.~\cite{bae2022noisy} proposed a noisy prediction calibration method based on a transition matrix to reduce the gap between noisy prediction and clean label based on a KNN prediction.

\noindent \textbf{Label-smoothing methods} rely on  modifying~\cite{szegedy2016rethinking,muller2019does} the sample-wise label distribution~\cite{lukasik2020does}.
Zhang et al.~\cite{zhang2021delving} propose online label smoothing (OLS) which generates soft labels by considering the relationships among multiple labels. 
Wei et al.~\cite{wei2021smooth} argue that the advantage of label smoothing vanishes under a high label noise regime since the label smoothing tends to over-smooth the estimated label classification, so they propose the generalised label smoothing (GLS)~\cite{wei2021smooth} which uses negative smoothing values for higher noise rates.
In general, label smoothing methods can underfit the training data since they tend to abandon the optimisation of hard clean-label samples~\cite{muller2019does}.

\noindent \textbf{Graph-based methods}
leverage the robustness of feature representations to discriminate between clean and noisy samples and regularise the training process.
Wu et al.~\cite{wu2020topological} explore the topological property of the data in the feature space to perform noise-cleaning by assuming that clean data are clustered together in this feature space, while the corrupted data are isolated.
Wu et al.~\cite{wu2021ngc} investigate the geometric structure of the data to model predictive confidence and filter out noisy samples. 
Iscen et al.~\cite{iscen2022learning} introduce a regularisation term that forces samples to have similar predictions to their neighbors. 
These graph-based methods have been designed 
for single-label classification, so they cannot be easily adapted to multi-label datasets.
Also, building a graph with multi-label data is also an issue for these methods.

\noindent \textbf{Multi-label Noisy label methods}
have received increasing attention in recent years due to the natural differences with respect to multi-class problems. Instead of the sample-wise noise found in multi-class problems, each label per sample can be corrupted in the multi-label scenario which can be problematic for selecting and correcting the label noise. In addition, the class imbalance~\cite{liu2021nvum} and the semantic divergence may also exacerbate the overfitting issue towards the majority classes. 
Zhao et al.~\cite{zhao2021evaluating} leverage label dependencies to handle noisy labels and use word embeddings to perform context-based regularization to avoid overfitting.
Li et al.~\cite{li2022estimating} consider the correlation between labels (i.e., ``fish'' and ``water'' have a stronger correlation when comparing with ``fish'' and ``sky'' ) to estimate the transition matrix. 
Xie et al.~\cite{xie2021partial} mitigate the negative impact of label noise by estimating the confidence for credible labels from the candidate label set.
Different from previous methods, we consider using the label semantic information and label smoothing techniques to capture more fine-grained image relationships and prevent the classifier from being overconfident on any of the noisy labels.

\subsection{Bag of Words}

The Bag of Words (BoW) method~\cite{harris1954distributional,sivic2003video,sivic2008efficient} is a traditional information retrieval technique, denoted by the representation of documents with a histogram of unordered words.
In computer vision, BoW~\cite{sivic2003video, csurka2004visual} represents images with a histogram of unordered local visual descriptors, learned from the training images in an unsupervised manner.
We adopt the BoW concept, but instead of extracting local visual descriptors (e.g., SIFT~\cite{lowe1999object}), we train a DNN to represent each image with a bag of global visual descriptors. 

    \begin{figure*}
        \centering
        \includegraphics[width=\linewidth]{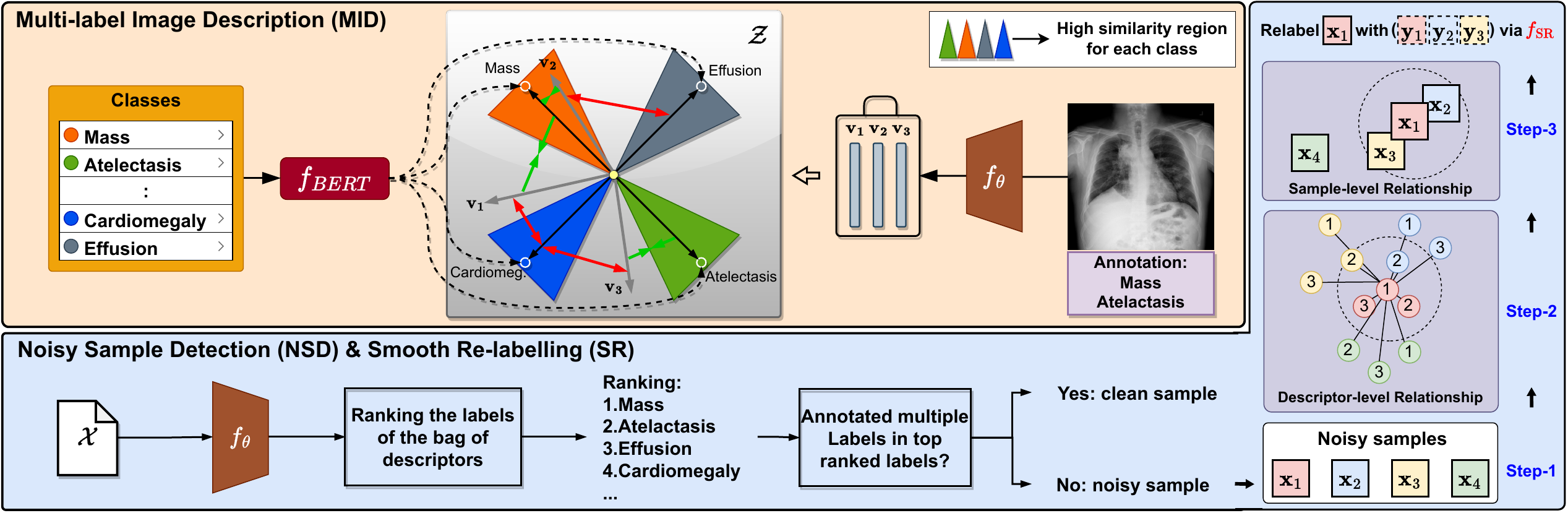}
        \vspace{-5pt}
        \caption{The \textbf{multi-label image description (MID)} module estimates a bag of visual descriptors $\{\mathbf{v}_m\}_{m=1}^{M}$ for each image, which is learned using the semantic information provided by BERT~\cite{peng2019transfer,johnson2016mimic,fiorini2018user}. 
        BERT represents each class with a descriptor (circle) in the semantic space $\mathcal{Z}$, where the triangular regions indicate high similarity regions of each corresponding class. 
        Please note that the \textcolor{green}{green} arrows represent the pulling of image descriptors towards one of the corresponding relevant embeddings in the rank loss of Eq.2. Conversely, the \textcolor{red}{red} arrows push the image descriptors away from all irrelevant embeddings and the rectangular regions indicate the high similarity area w.r.t to each class's word embeddings.
        The \textbf{noisy sample detection (NSD)} module leverages the consistency between label ranking prediction from MID and the original annotation to detect the noisy samples, which are then \textbf{smoothly re-labelled (SR)} with $f_{\text{SR}}(.)$ in~\eqref{eq:mixup}. 
        }
        \label{fig:mid}
        \vspace{-10pt}
    \end{figure*}
\section{Method} \label{sec:method}

    We assume the availability of a noisy multi-label training set denoted by $\mathcal{D}=\{\mathbf{x}_i, \mathbf{y}_i\}_{i=1}^{|\mathcal{D}|}$, where $\mathbf{x}_i\in\mathcal{X}\subset \mathbb{R}^{H \times W \times R}$ represents an image of size $H \times W$ and $R$ colour channels\footnote{We consider each image to have 3 colour channels (R=3) and dimensions of $H=W=512$ for NIH, and $H=W=224$ for CXP.}, and $\mathbf{y}_i \in \mathcal{Y} = \{0, 1\}^{|\mathcal{Y}|}$ denotes the multi-label annotation. The testing set is similarly defined. 
    
    \subsection{Bag of Multi-label Descriptors (BoMD)}
    Our method, described in Alg.~\ref{alg:BoMD}, is inspired by the observation that a noisy-labelled sample tends to be an outlier surrounded by clean-labelled samples in the feature space~\cite{wu2020topological}. 
    Hence, the label of each sample should be consistent with the labels of the neighboring samples.
    This motivated us to develop an approach to re-label a noisy multi-label image with the estimated label distribution from its neighbourhood.
    The proposed BoMD has two stages (\cref{fig:motivation}): 1) image description learning that transforms a training image into a bag of visual descriptors that lie in the semantic space $\mathcal{Z}\subset \mathbb{R}^{Z}$ populated by word embeddings computed from image labels~\cite{peng2019transfer} 
    2) graph construction to smoothly re-label noisy multi-label images, where each image is represented by a sub-graph built from the learned bag of visual descriptors, which can capture fine-grained image relationships. This smoothly re-labelled dataset is then used for training a multi-label classifier.

    \subsection{Multi-label Image Description (MID)}
    Motivated by BoW, our MID (\cref{fig:mid}) represents an image by associating its multiple labels to a bag of global visual descriptors.
    MID projects the image into BERT's semantic space using a set of visual descriptors that are optimised to promote their similarity with the semantic descriptors produced by BERT models from the multi-label annotation of the image.
    The MID of image $\mathbf{x}$ are extracted with 
    $\mathcal{V} = f_{\theta}(\mathbf{x})$, where $\mathcal{V} = \{ \mathbf{v}(m) \}_{m=1}^{M}$ denotes the $M$\footnote{We empirically set $M=3$ according the ablation study of the hyper-parameters in Supp. Material.} visual descriptors in BERT's semantic space, i.e., $\mathbf{v}(m) \in \mathcal{Z}$.
    The BERT language models (e.g., BlueBERT~\cite{peng2019transfer}, medical language model pre-trained on PubMed abstracts~\cite{fiorini2018user} and clinical notes~\cite{johnson2016mimic})  produce semantic descriptors in the form of word embeddings with $\mathbf{w}(c) = f^{\mathcal{Y}}_{BERT}(c)$ for $c \in \{1,...,|\mathcal{Y}|\}$, forming $\mathcal{W} = \{\mathbf{w}(c)\}_{c=1}^{|\mathcal{Y}|}$, where $\mathbf{w}(c) \in \mathcal{Z}$, with $\mathcal{Z}$ being the same space as for $f_{\theta}(.)$.  
    More specifically, MID is trained with:
    \begin{equation}
        \theta^{*} = \arg\min_{\theta}  \frac{1}{|\mathcal{D}|}\sum_{(\mathbf{x}_i, \mathbf{y}_i)\in\mathcal{D}} \omega_i \ell_{mid}(\mathbf{x}_i, \mathbf{y}_i,\theta) + \beta\ell_{reg}(\mathbf{x}_i,\theta)
        \label{eq:optimise_theta}
    \end{equation}
    where \scalebox{0.9}{$\omega_i = \left(\left(\sum_{c=1}^{|\mathcal{C}|}\mathbb{I}(\mathbf{y}_i(c)=1)\right) \div \left( \sum_{c=1}^{|\mathcal{C}|}\mathbb{I}(\mathbf{y}_i(c)=0)\right)\right)$} is a normalisation that controls the ranking weight based on the number of positive and negative labels ($\mathbb{I}(.)$ represents an indicator function)~\cite{burges2005learning,zhang2016fast}, and the hyper-parameter $\beta$ weights the regulariser. Also  in~\eqref{eq:optimise_theta} we have:
    \begin{equation}
    \scalebox{0.9}{$
    \begin{aligned}
    \ell_{mid}&(\mathbf{x}_i, \mathbf{y}_i,\theta) = \\
    &\sum_{\substack{p=1, \\ \mathbf{y}_i(p)=1}}^{|\mathcal{Y}|}
    \sum_{\substack{n=1, \\ \mathbf{y}_i(n)=0}}^{|\mathcal{Y}|} \log{\left(1+\exp{(\ell_{rank}(\mathcal{V}_i, \mathcal{W}, p, n))}\right)}, \text{ where}\\
    \ell_{rank}&(\mathcal{V}_i, \mathcal{W}, p, n)  = \left(\max_{\mathbf{v}\in\mathcal{V}_i}(\langle \mathbf{v}, \mathbf{w}(n)\rangle) - \max_{\mathbf{v}\in\mathcal{V}_i}(\langle \mathbf{v}, \mathbf{w}(p) \rangle)\right),
    \end{aligned}
    $}
    \label{eq:ell_val}
    \end{equation}
    with $\langle.,.\rangle$ representing the dot product operator, $p,n \in \{1,...,|\mathcal{Y}|\}$ denoting the indices to the positive (i.e., $\mathbf{w}(p)$ where $\mathbf{y}_i(p)=1$) and negative word embeddings (i.e., $\mathbf{w}(n)$ where $\mathbf{y}_i(n)=0$), respectively,
    $\mathcal{V}_i = f_{\theta}(\mathbf{x}_i)$, and
    \begin{equation}
        \ell_{reg}(\mathbf{x}_i,\theta)=
    \sum_{\mathbf{v}(m) \in \mathcal{V}_i} \frac{(\mathbf{v}(m) - \bar{\mathbf{v}}(m))^{\top}(\mathbf{v}(m) - \bar{\mathbf{v}}(m))}{Z-1}
    \label{eq:ell_reg}
    \end{equation}
    being a regulariser to reduce descriptor variance,
    where $Z$ is the number of dimensions of $\mathcal{Z}$, and $\bar{\mathbf{v}}(m)$ denoting the MID mean in $\mathcal{V}_i$.
    The $\ell_{mid}(.)$  in~\eqref{eq:ell_val}, inspired by previous multi-label learning methods~\cite{zhang2016fast}, forces the dot product $\langle \mathbf{v}, \mathbf{w}(p) \rangle$ to rank higher than the $\langle \mathbf{v}, \mathbf{w}(n) \rangle$, which means that the visual descriptors will be more similar to positive label embeddings than the negative label ones.
    Intuitively, this loss encourages semantically similar image descriptors to cluster around their related semantic descriptors, which will benefit our graph-based smooth re-labelling.

    \subsection{Graph Construction and Smooth Re-labelling}
    Considering that the learned visual descriptors are likely to be closer to clean labels in the semantic space, we formulate the detection of noisy training samples by first ranking (in descending order of similarity) the labels for image $\mathbf{x}_i$ (according to the inner product with the word embeddings from the labels), as follows:
    \begin{equation}
        \mathbf{r}_i = \arg\text{sort}_{c \in \{1,...,|\mathcal{Y}|\}}\left [ \max_{\mathbf{v}\in\mathcal{V}_i}(\langle \mathbf{v},\mathbf{w}(c) \rangle) \right ],
        \label{eq:rank_clean_noisy}
    \end{equation}
    where $\mathbf{r}_i(1) \in \{1,...,|\mathcal{Y}|\}$ is the highest ranked label and $\mathbf{r}_i(|\mathcal{Y}|)  \in \{1,...,|\mathcal{Y}|\}$ is the lowest ranked label. Then, clean samples are the ones where $\mathbf{r}_i(p) < \mathbf{r}_i(n)$ for all positive labels $p \in \mathcal{P}_i$ and negative labels $n \in \mathcal{N}_i$, with $\mathcal{P}_i = \{c | \mathbf{y}_i(c)=1\}$ and $\mathcal{N}_i = \{c | \mathbf{y}_i(c)=0\}$; otherwise, the sample is classified as noisy.

    The second stage of BoMD re-labels noisy samples using the sample graph built with the MID visual descriptors. The graph is constructed by representing each training image $\mathbf{x}_i$ with $M$ nodes $\{\mathbf{v}_i(m)\}_{m=1}^{M}$ from $\mathcal{V}_i=f_{\theta}(\mathbf{x}_i)$, where the edge weight between the $m^{th}$ descriptor of the $i^{th}$ image and the $n^{th}$ descriptor of the $j^{th}$ image is defined by 
    $e(\mathbf{v}_i(m),\mathbf{v}_j(n)) = 1/\| \mathbf{v}_i(m)-\mathbf{v}_j(n) \|_2$.
    This means that the graph has the set of nodes denoted by $\{ \mathbf{v}_i(m) \}_{i=1,m=1}^{|\mathcal{D}|,M}$ and edges $\{ e(\mathbf{v}_i(m),\mathbf{v}_j(n)) \}_{i,j=1,m,n=1}^{|\mathcal{D}|,M}$.
    The re-labelling is based on finding the $K$ nearest neighbouring training images to image $i$ by using the graph nodes, with:
    \begin{equation}
    \scalebox{0.9}{$
        \mathcal{K}(\mathcal{V}_i) = \text{top}K_{j \in \{1,...,|\mathcal{D}|\},
        m,n \in \{1,...,M\}}(e(\mathbf{v}_i(m),\mathbf{v}_j(n))),
        $}
        \label{eq:knn}
    \end{equation}
    where $\mathcal{K}(\mathcal{V}_i)$ contains the unique image indices from the $K$ nodes with the largest edge weights. 
    Next, for all samples identified as noisy, we update their labels with $\mathbf{\tilde{y}}_i=f_{\text{SR}}(\mathbf{y}_i,\mathbf{\bar{y}}_i)$, which is defined as
    \begin{equation}
        \scalebox{0.95}{$
        f_{\text{SR}}(\mathbf{y}_i,\mathbf{\bar{y}}_i) = (1-\lambda) \cdot \mathbf{y}_i + \lambda \cdot \left(\gamma \cdot \mathbf{1}_{|\mathcal{Y}|} + (1-\gamma)\cdot \bar{\mathbf{y}}_i\right) \odot \mathbf{m},
        $}
        \label{eq:mixup}
    \end{equation}
    where 
    $\lambda\in[0, 1], \gamma \in [0,0.5]$,
    $\bar{\mathbf{y}}_i = \frac{1}{K}\sum_{j \in \mathcal{K}(\mathcal{V}_i)}\mathbf{y}_j$, 
    $\mathbf{1}_{|\mathcal{Y}|}$ denotes a vector with ones of size $|\mathcal{Y}|$ (uniform distribution) to prevent the re-labelling from being overconfident on any of the labels,
    $\odot$ is the element-wise vector multiplication, and
    $\mathbf{m} = \mathbb{I}((\mathbf{y}_i+\bar{\mathbf{y}}_i) > 0)$ 
    is a binary mask to filter out high confident negative labels (with $\mathbb{I}(.)$ being the indicator function) to mitigate the over-smoothing issue.

\begin{algorithm}[t]
\caption{BoMD}
\label{alg:BoMD}
\definecolor{codeblue}{rgb}{0.25,0.5,0.5}
\definecolor{codekw}{rgb}{0.85, 0.18, 0.50}
\definecolor{ao}{rgb}{0.0, 0.5, 0.0}
\begin{algorithmic}[1]
    \State \textbf{require:} Training set $\mathcal{D}$, and BERT model embeddings $\mathbf{w}(c) = f^{\mathcal{Y}}_{BERT}(c)$ for $c \in \{1,...,|\mathcal{Y}|\}$
    \State \textcolor{ao}{\# Build MID}
    \State Train $f_\theta(.)$ with~\eqref{eq:optimise_theta} using $f_{BERT}^{\mathcal{Y}}(.)$ and $\mathcal{D}$
    \State Create re-labeled set $\tilde{\mathcal{D}}=\emptyset$ and
    noisy set $\mathcal{B}=\emptyset$
    \State \textcolor{ao}{\# Detect noisy samples}
    \For{$(\mathbf{x}_i,\mathbf{y}_i) \in \mathcal{D}$}
    \State Compute label rank $\mathbf{r}_i$ from~\eqref{eq:rank_clean_noisy} 
    \State \textbf{If} $\mathbf{r}_i(p) > \mathbf{r}_i(n)$ for all $p \in \mathcal{P}_i$ and all $n \in \mathcal{N}_i$
    \State \textbf{Then} $\tilde{\mathcal{D}} \leftarrow \tilde{\mathcal{D}} \bigcup (\mathbf{x}_i,\mathbf{y}_i)$ 
    \State \textbf{Else} $\mathcal{B} \leftarrow \mathcal{B} \bigcup (\mathbf{x}_i,\mathbf{y}_i)$
    \EndFor
    \State Build graph with nodes $\{ \mathbf{v}_i(m) \}_{i=1,m=1}^{|\mathcal{D}|,M}$ and edges $\{ e(\mathbf{v}_i(m),\mathbf{v}_j(n)) \}_{i,j=1,m,n=1}^{|\mathcal{D}|,M}$ using $f_{\theta}(.)$ and $\mathcal{D}$
    \State \textcolor{ao}{\# Re-label noisy samples}
    \For{$(\mathbf{x}_i,\mathbf{y}_i) \in \mathcal{B}$}
    \State Re-label $\mathbf{x}_i$ with $\tilde{\mathbf{y}}_i$ from~\eqref{eq:mixup}
    \State $\tilde{\mathcal{D}} \leftarrow \tilde{\mathcal{D}} \bigcup (\mathbf{x}_i,\tilde{\mathbf{y}}_i)$
    \EndFor
    \State \textcolor{ao}{\# Train final classifier}
    \State Train $f_{\phi}:\mathcal{X} \to [0,1]^{|\mathcal{Y}|}$ with BCE loss using $\tilde{\mathcal{D}}$
\end{algorithmic}
\end{algorithm}

\subsection{Training and Testing}

We build a new training set $\tilde{\mathcal{D}} = \{ (\mathbf{x}_i,\tilde{\mathbf{y}}_i) | (\mathbf{x}_i, \mathbf{y}_i) \in \mathcal{D}\}_{i=1}^{|\mathcal{D}|}$, where $\tilde{\mathbf{y}}_i = \mathbf{y}_i$ if sample $(\mathbf{x}_i, \mathbf{y}_i)$ is clean from~\cref{eq:rank_clean_noisy}, or computed from~\cref{eq:mixup} if sample is noisy$\}$.
Then, we train a regular classifier $f_{\phi}:\mathcal{X} \to [0,1]^{|\mathcal{Y}|}$ by minimizing a BCE loss on $\tilde{\mathcal{D}}$.  Testing is based on applying the trained $f_{\phi}(.)$ to test images.

\section{Experiments}
    Our experiments are based on the  following datasets.
    \noindent \textbf{Noisy Training Sets.} The \textbf{\textit{NIH Chest X-ray14} (NIH)}~\cite{nih}  contains 112,120 frontal-view CXR images from 30,805 patients, where each image has between 0 and 14 annotated pathologies and the training set contains 86,524 images with a maximum of 9 labels per image. 
    The \textbf{\textit{CheXpert (CXP)}}~\cite{cxp} has 224,316 frontal-view CXR images from 65,240 patients labelled with 14 common chest radiographic observations, where the training set contains 170,958 images with a maximum of 8 labels per image. The labels of these two datasets are obtained from an NLP algorithm, which forms noisy multi-label annotations~\cite{oakden2017exploring}.
    
    \noindent \textbf{Clean Testing Sets.} The \textbf{\textit{OpenI}}~\cite{openi} dataset contains 3,999 radiology reports and 7,470 frontal/lateral-view CXR images from the Indiana Network for Patient Care. We use all frontal-views images for evaluation, resulting in 3,818 images and 19 manually annotated diseases. 
    We also use  \textbf{\textit{PadChest}}~\cite{padchest}, which contains 160,861 images with 27 chest radiographic observations. PadChest has a mixture of machine and manually labelled images, but we only use the manually labelled frontal-view images (about 15.25\%\footnote{The $27\%$ of PadChest's annotated images include the lateral-view CXRs.} of the images).
    Additionally, we follow previous works~\cite{xue2022robust, liu2021nvum} to evaluate the model on a re-organised subset of the official NIH test set~\cite{nih}, referred to as \textbf{\textit{NIH-GOOGLE}}, with 1,962 CXR images that focuses on two findings (pneumothorax and nodule/mass). Each image is manually re-labelled by at least three 
    certified radiologists~\cite{majkowska2020chest} with final label pooled from an adjudication process.

    \noindent \textbf{Systematic Noisy-label Assessment}. 
    We introduce the first systematic noisy multi-label assessment benchmark by combining the NIH~\cite{nih} dataset and the clean test samples from Padchest~\cite{padchest} as a new noisy-label training set with the OpenI~\cite{openi} as the clean testing set. 
    We then apply symmetric label noise~\cite{li2020dividemix, bae2022noisy} to the PadChest training subset \textbf{only}, where we flip the labels from present to absent, and vice-versa, based on two control variables, namely: 1) the proportion of noisy samples, and 2) the probability of switching a label. This dataset is referred to as \textbf{\textit{NIHxPDC}}.

    \subsection{Implementation Details}
    
    We resize the NIH~\cite{nih} images to $512\times512$, and CXP~\cite{cxp} images to $224\times224$, where images are normalised using ImageNet~\cite{russakovsky2015imagenet} mean and standard deviation. We use random resize crop and random horizontal flipping as data augmentation. 
    The BlueBERT word embeddings in $\mathcal{W}$ are L2 normalised. For the MID model $f_{\theta}(.)$, we use the ImageNet~\cite{russakovsky2015imagenet} pre-trained DenseNet121~\cite{huang2017densely}, which is trained with Adam optimiser~\cite{kingma2014adam} using a learning rate of 0.0001 with cosine annealing decay~\cite{loshchilov2016sgdr}, batch size of 16 and 20 epochs. We set the number of descriptors $M=3$ and weight of regulariser $\beta=0.3$.
    The descriptor graph is implemented with Faiss~\cite{johnson2019billion} for efficient search (the search process in~\cref{eq:knn} takes 5 seconds for all training samples from NIH).
    The classifier $f_{\phi}(.)$ uses another ImageNet pre-trained DenseNet121~\cite{huang2017densely}, and then it is trained with Adam optimiser~\cite{kingma2014adam} using a learning rate of 0.05, batch size of 16 and 30 epochs. 
    We empirically set the mixup coefficient $\lambda$ and $\gamma$ in~\cref{eq:mixup} to 0.6 and 0.25, and use $K=10$ in~\cref{eq:knn}.
    The classification results are assessed with the mean of the class-wise area under the receiver operating characteristic curve (AUC) for all disease classes~\cite{cxp,hermoza2020region, liu2021acpl}.
    All experiments are implemented with Pytorch~\cite{paszke2019pytorch} and run on an NVIDIA RTX3090 GPU (24GB). 
    Training takes 23h for NIH and 15h for CheXpert, and  
    testing for a single image takes 13.41ms for NIH and 12.24ms for CheXpert.
    
    \begin{table}[!t]
        \centering
        \resizebox{0.95\linewidth}{!}{  
        \begin{tabular}{c|ccc}
        \toprule						
        Methods	&	Models	&	OpenI	&	PadChest	\\	\midrule \midrule																			
        \multirow{2}{*}{General}	&	Hermoza et al~\cite{hermoza2020region}	&	85.54 ± 0.42	&	83.90 ± 0.57	\\																				
        ~	&	CAN~\cite{ma2019multi}	&	84.26 ± 0.35	&	83.10 ± 0.25	\\	\midrule																			
        \multirow{2}{*}{Noise-cleaning}	&	DivideMix~\cite{li2020dividemix}	&	72.76 ± 1.09	&	75.49 ± 0.21	\\																				
        ~	&	FINE~\cite{kim2021fine}	&	63.67 ± 1.78 	& 70.91	± 0.20	\\	\midrule																			
        \multirow{2}{*}{Noise-robust}	&	ELR~\cite{liu2020early}	&	86.62 ± 0.87	&	85.24 ± 0.11	\\																				
        ~	&	\blueb{NVUM}~\cite{liu2021nvum}	&	\blueb{88.17 ± 0.48}	&	\blueb{85.49 ± 0.06}	\\	\midrule			\multirow{1}{*}{Transition matrix} 	&	NPC~\cite{bae2022noisy}	&	86.21 ± 0.07	&	83.88 ± 0.05	\\	\midrule
        \multirow{1}{*}{Graph-based} 	&	NCR~\cite{iscen2022learning}	&	85.06 ± 0.96	&	83.79 ± 0.48	\\	\midrule
        \multirow{5}{*}{Label smoothing}															
        ~	&	LS~\cite{lukasik2020does}	&	83.72 ± 1.29	&	80.93 ± 0.82	\\																				
        ~	&	OLS~\cite{zhang2021delving}	&	85.08 ± 0.31	&	83.51 ± 0.61	\\																				
        ~	&	GLS~\cite{wei2021smooth}	&	83.80 ± 0.34	&	81.56 ± 0.24	\\																				
        ~	&	\redb{BoMD}	&	\redb{89.57 ± 0.22}	&	\redb{86.45 ± 0.08}	\\				
        \bottomrule								
        \end{tabular}
        }
        \vspace{-5pt}
        \caption{Mean $\pm$ standard deviation AUC results for the testing sets from OpenI and PadChest, using models \textbf{trained on NIH~\cite{nih}}. Best and the second best results are in \redb{red}/\blueb{blue}. }
        \label{tab:nih_result}
        \vspace{-12pt}
    \end{table}

    \subsection{Classification Results on Real-world Datasets}
    
    We first compare the performance of state-of-the-art (SOTA) methods with our BoMD in \cref{tab:nih_result} and \cref{tab:cxp_result}.
    We run each experiment three times and show mean and standard deviation of AUC results.
    \cref{tab:nih_result} shows the testing AUC results of the training with NIH~\cite{nih} and testing on OpenI~\cite{openi} and PadChest~\cite{padchest}. 
    Our model surpasses the second best method~\cite{liu2021nvum} by 1.4\% and 0.96\% on the two test sets with $p$-values $0.0018$ and $10^{-14}$, respectively (one-sided z-test).
    We also report testing performance on OpenI~\cite{openi} and PadChest~\cite{padchest} using training on CXP~\cite{cxp} in~\cref{tab:cxp_result}. 
    Our model surpasses the second best result~\cite{cxp} by 2.82\% and 1.13\% on the two test sets with $p$-values $0.0002$ and $10^{-14}$, respectively (one-sided z-test).
    We also notice that there is a large gap between all models' performance for certain classes. 
    For example, in our model, the AUC results for \textit{Pneumonia} classification when training on NIH are much better than when training on CXP, with a gap of 23.20\% and 15.48\% on OpenI~\cite{openi} and PadChest~\cite{padchest}, respectively, which may be due to CXP's smaller image size. 
    The NIH-GOOGLE~\cite{xue2022robust,liu2021nvum} evaluation on classes pneumothorax and nodule/mass (obtained from the average classification scores for Mass and Nodule) is displayed in~\cref{tab:nih_google}, which shows that our method outperforms the SOTA methods on both Pneumothorax (+0.6\% compared to~\cite{xue2022robust}) and Mass/Nodule (+2.4\% compared~\cite{liu2021nvum}) classifications. The per-finding results are reported in the \textit{Supplementary Material}. 

    \begin{table}[!t]
        \centering
    
        \resizebox{0.95\linewidth}{!}{  
        \begin{tabular}{c|ccc}
        \toprule								
        Methods	&	Models	&	OpenI	&	PadChest	\\	\midrule \midrule
        \multirow{2}{*}{General}	&	Hermoza et al~\cite{hermoza2020region}	&	74.94 ± 0.50	&	77.24 ± 0.04	\\	
        ~	&	CAN~\cite{ma2019multi}	&	76.34 ± 0.49	&	78.92 ± 0.58 	\\	\midrule
        \multirow{2}{*}{Noise-cleaning}	&	DivideMix~\cite{li2020dividemix}	&	73.23 ± 0.60	&	74.21 ± 0.55	\\	
        ~	&	FINE~\cite{kim2021fine}	&	71.68 ± 0.54	&	73.83 ± 0.52	\\	\midrule
        \multirow{2}{*}{Noise-robust}	&	ELR~\cite{liu2020early}	&	77.16 ± 0.79	&	79.97 ± 0.71	\\
        ~   &	\blueb{NVUM}~\cite{liu2021nvum}	&	\blueb{77.21 ± 0.81}	&	\blueb{80.62 ± 0.10}	\\	\midrule
        \multirow{1}{*}{Transition matrix}	&	NPC~\cite{bae2022noisy}	&	75.32 ± 0.40	&	77.30 ± 0.03	\\	\midrule
        \multirow{1}{*}{Graph-based} & NCR~\cite{iscen2022learning}	&	76.93 ± 0.38	&	79.36 ± 0.84	\\	\midrule
        \multirow{4}{*}{Label Smoothing}	&	LS~\cite{lukasik2020does}	&	72.86 ± 0.23	&	75.34 ± 0.50	\\	
        ~	&	OLS~\cite{zhang2021delving}	&	76.52 ± 0.83	&	77.72 ± 0.70	\\	
        ~	&	GLS~\cite{wei2021smooth}	&	76.50 ± 0.26	&	78.80 ± 0.70	\\	
        ~	&	\redb{BoMD}	&	\redb{80.03 ± 0.73}	&	\redb{81.76 ± 0.40}	\\	
        \bottomrule														
        \end{tabular}
        }
        \vspace{-5pt}
        \caption{Mean $\pm$ standard deviation AUC results for the testing sets from OpenI and PadChest, using models \textbf{trained on CXP~\cite{cxp}}. Best and the second best results are in \redb{red}/\blueb{blue}.}
        \label{tab:cxp_result}
        \vspace{-12pt}
    \end{table}
    
    \begin{table*}[t]
    \centering
    \resizebox{0.9\textwidth}{!}{%
    \begin{tabular}{l|cccccccc|c}
    \toprule \hline
            & BCE  & F-correction~\cite{patrini2017making} & MentorNet~\cite{jiang2018mentornet} & Decoupling~\cite{malach2017decoupling} & Co-teaching~\cite{han2018co} & ELR~\cite{liu2020early}  & Xue et al.~\cite{xue2022robust} & NVUM~\cite{liu2021nvum} & BoMD \\ \hline
    Pneu   & 87.0 & 80.8         & 86.6      & 80.1       & 87.3        & 87.1 & \blueb{89.1}       & 88.9 & \redb{89.7}\\ 
    M/N    & 84.3 & 84.8         & 83.7      & 84.3       & 82.0        & 83.2 & 84.6       & \blueb{85.5} & \redb{87.9} \\ \hline \bottomrule
    \end{tabular}%
    }
    \vspace{-5pt}
    \caption{Pneumothorax and Mass/Nodule AUC of NIH-Google~\cite{majkowska2020chest} for models trained on NIH~\cite{nih}. Best and the second best results for OpenI and PadChest are in \redb{red}/\blueb{blue}.}
    \label{tab:nih_google}
    \vspace{-10pt}
    \end{table*}
    
    \subsection{Systematic Noisy-label Benchmark}

    In our systematic noisy-label benchmark NIHxPDC, we compare our BoMD with the SOTA method NVUM~\cite{liu2021nvum} and a baseline model trained with BCE loss. Recall that this benchmark relies on two control variables: a) percentage of noisy samples $p_s$, and b) probability of switching a label $p_l$, where $p_s,p_l \in \{0\%,20\%,40\%,60\%\}$.
    We show the mean AUC classification over the 14 classes on clean OpenI in Tab.~\ref{tab:noise_auc}. 
    Notice that our BoMD has better results than NVUM and BCE for the majority of the cases, with a 3\% to 5\% improvement compared with NVUM and BCE when $p_s=20\%$, 1\% to 4\% improvement for $p_s=40\%$, and 1\% to 5\% improvement for $p_s=60\%,p_l=20\%,40\%$, but for $p_s,p_l=60\%$, NVUM improves $2.8\%$ over BoMD.
    Hence, BoMD provides a consistently better classification result across many noise rates, but for large noise rates, our re-labelling method may be injecting too much noise into the training set. Another interesting point to note is that with $0\%$ controlled noise, BoMD shows $89.76\%$ AUC, which means that the AUC drops an average of $3.8\%$ for each addition of $20\%$ for $p_s$ with a fixed $p_l = 20\%$.

    \begin{table}[t]
        \centering
        \resizebox{\linewidth}{!}{%
        \begin{tabular}{c|c|ccc|ccc|ccc}
        \toprule \hline
        $p_s$ & \multicolumn{1}{c|}{0$\%$} & \multicolumn{3}{c|}{20$\%$} & \multicolumn{3}{c|}{40$\%$} & \multicolumn{3}{c}{60$\%$}  \\ \hline
        $p_l$ & 0\% & 20$\%$ & 40$\%$ & 60$\%$ & 20$\%$ & 40$\%$ & 60$\%$ & 20$\%$ & 40$\%$ & 60$\%$ \\ \hline
        BCE  & 85.65 & 83.99 & 81.42 & 79.63 & 80.99 & 78.51 & 75.79 & 77.14 & 75.35 & 72.39 \\ \hline
        NVUM & 87.89 & 85.34 & 82.83 & 81.35 & 82.52 & 80.64 & 78.66 & 78.49 & 77.19 & \redb{76.91} \\ \hline
        BoMD & \redb{89.76} & \redb{88.00} & \redb{86.26} & \redb{84.55} & \redb{84.47} & \redb{81.86} & \redb{78.68} & \redb{82.23} & \redb{78.15} & 74.11 \\ \hline \bottomrule
        \end{tabular}
        }
        \vspace{-5pt}
        \caption{Mean testing AUC results for the 13 OpenI classes with models trained on NIHxPDC. Best results  in \redb{red}. 
        }
        \label{tab:noise_auc}
        \vspace{-10pt}
    \end{table}

    \begin{figure}[ht]
        \centering
        \subfloat[\centering Recall]{{\includegraphics[width=0.455\linewidth]{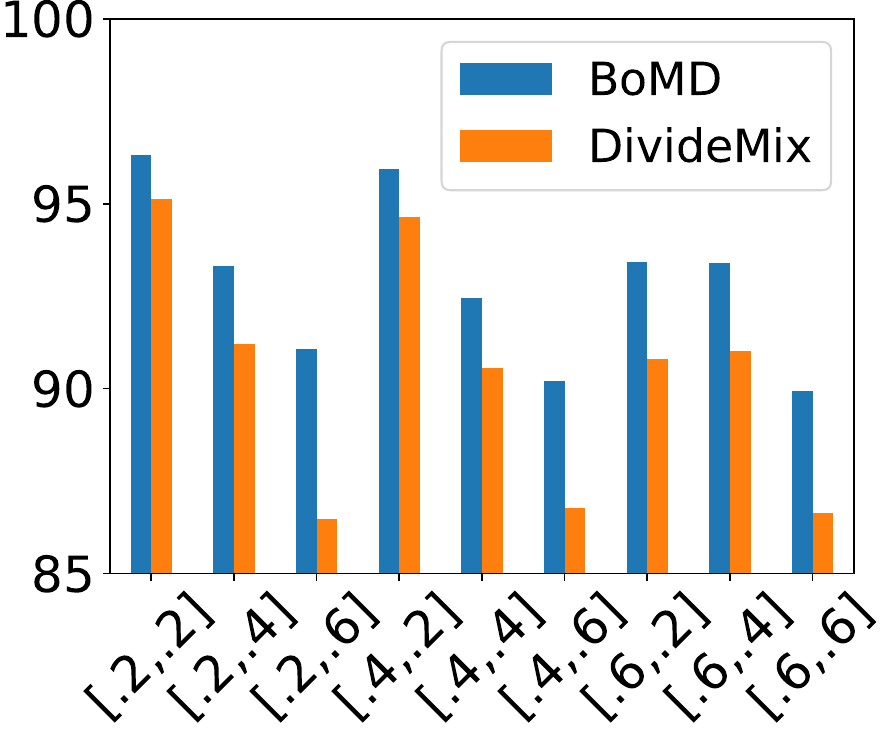} }}%
        \subfloat[\centering Precision]{{\includegraphics[width=0.455\linewidth]{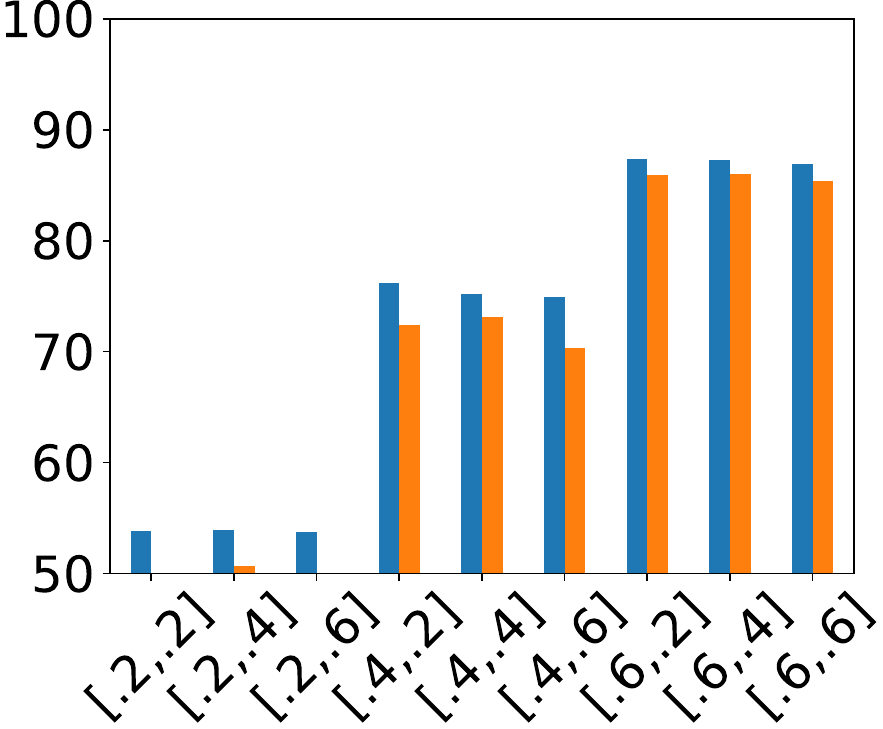} }}%
        \vspace{-5pt}
        \caption{Noisy-label sample detection performance on NIHxPDC. We compare our proposed  rank-based detection approach with DivideMix's small loss method~\cite{li2020dividemix} for a) recall, and b) precision. The horizontal axes show the values for $[p_s,p_l]$.}
        \vspace{-5pt}
        \label{fig:noise_det}
    \end{figure}
    
    We now study the effectiveness of the detection and re-labelling of noisy samples by BoMD. 
    In Fig.~\ref{fig:noise_det}, we compare the precision and recall of our detection of noisy-label samples compared with the traditional small-loss approach used by  DivideMix~\cite{li2020dividemix}. Notice that our noisy-label sample detection consistently outperforms DivideMix's small-loss method on both measures. 
    An interesting note is that while recall worsens, precision improves with increasing noise rates.  
    This happens because of the natural imbalance found in the distribution of classes in CXR datasets, where the class "No Findings" is dominant. A larger synthetic noise rate implies that this class is more affected than the others, but at the same time easier to detect given that the NIH dataset has relatively smaller noise rates.

    \begin{figure}[t]
        \centering
        \vspace{-10pt}
        \subfloat[\centering Clean]{{\includegraphics[width=0.32\linewidth]{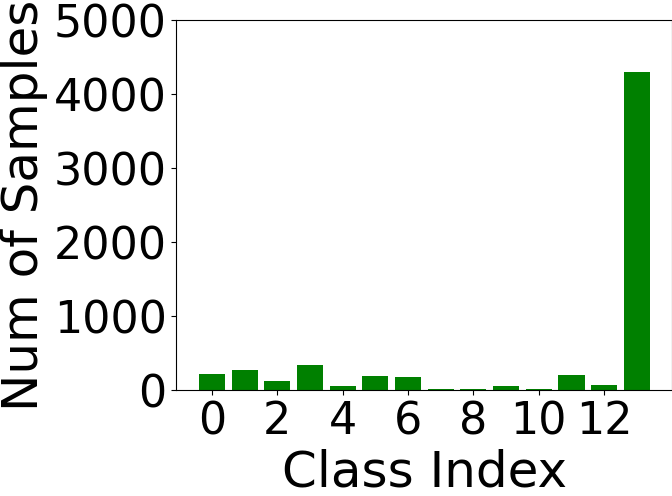} }}%
        \subfloat[\centering Noise]{{\includegraphics[width=0.32\linewidth]{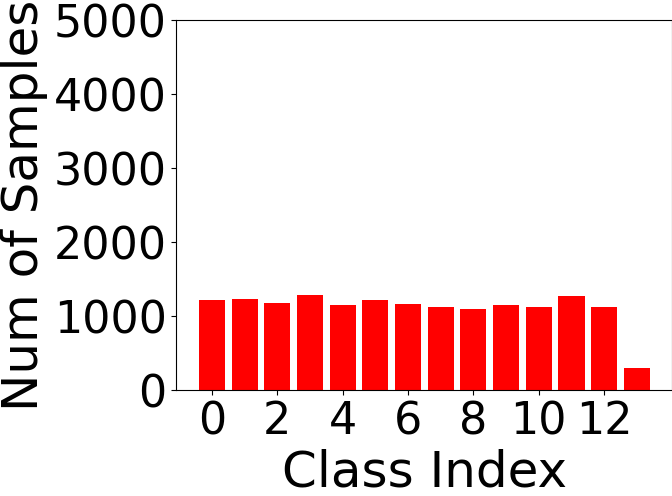} }}%
        \subfloat[\centering Re-labelled]{{\includegraphics[width=0.32\linewidth]{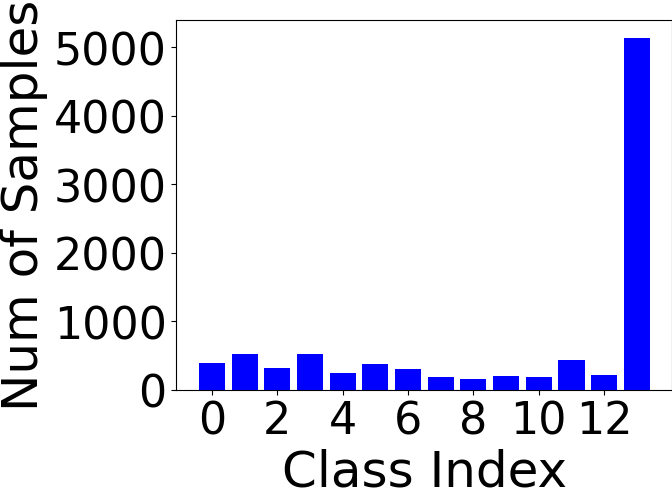} }}%
        \vspace{-5pt}
        \caption{Visualisation of the changes in the histogram of label distributions after applying our re-labelling.\textbf{Left}: label distribution for the clean set OpenI. \textbf{Middle}: label distribution after injecting symmetric noise $p_s=0.4,p_l=0.2$. \textbf{Right}: label distribution after the re-labelling by BoMD.}
        \label{fig:label_dist_shift}
        \vspace{-10pt}
    \end{figure}
    
\begin{figure}[t]
        \centering       
        \vspace{-2pt}
        \includegraphics[width=0.85\linewidth]{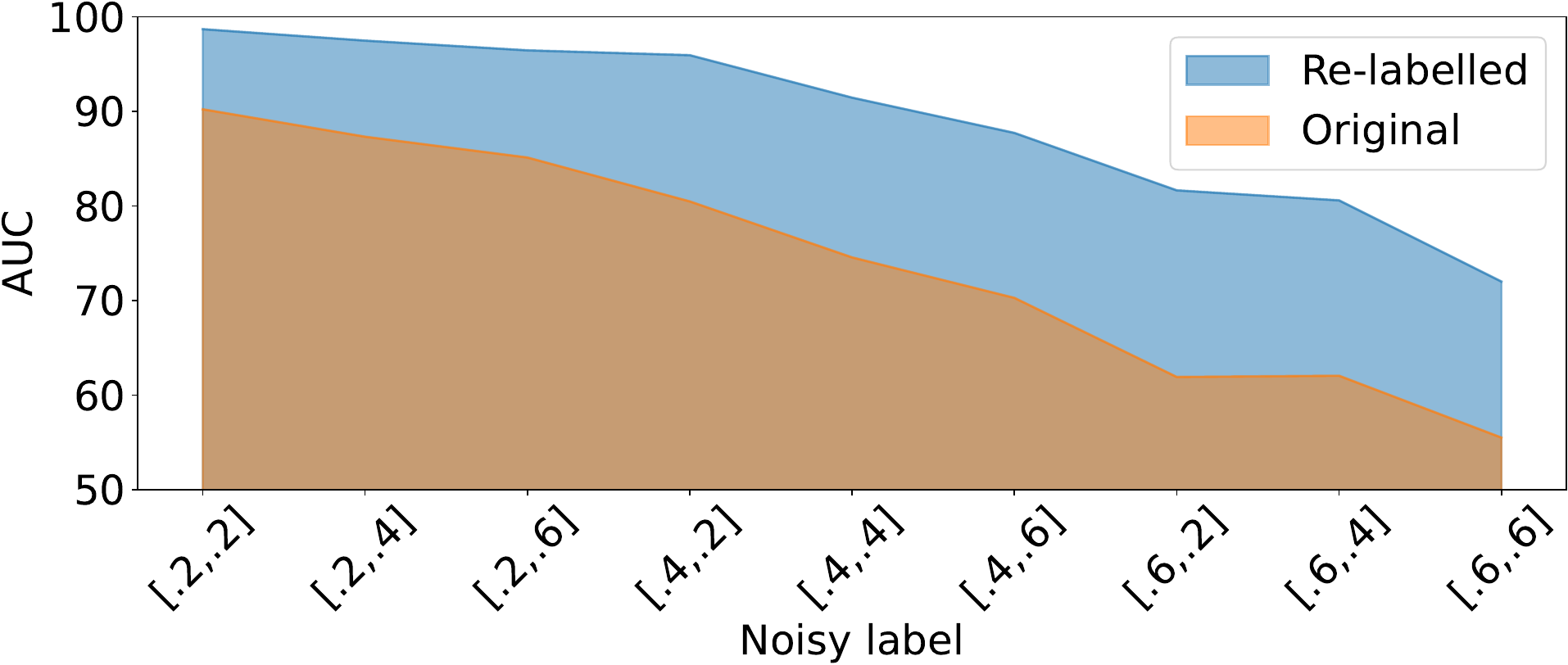}
        \vspace{-5pt}
        \caption{Mean AUC over labels before (orange) and after (blue) our re-labelling w.r.t PadChest's clean labels. The horizontal axes show values for $[p_s,p_l]$.}
        \label{fig:relabel_auc}
        \vspace{-15pt}
    \end{figure}

    We also visualise in Fig.~\ref{fig:label_dist_shift} the label distribution change before and after applying our re-labelling. 
    Note that our method successfully corrects the noisy label distribution to be closer to the original clean label distribution.
    The mean AUC over the labels in the re-labelled dataset before and after our re-labelling process is presented in~\cref{fig:relabel_auc}, where results show that our re-labelling process significantly improves the label cleanliness of the training set for all benchmark noise rates.
    Recall that in multi-class problems, such re-labelling is facilitated by the fact that each sample can only have a single label.
    However, such constraint is dropped for multi-label problems, making the re-labelling more complicated because the feature space will be populated with multiple clusters containing different combinations of multi-labels.
    In the supplementary material, 
    we evaluate the amount of consistency the KNN neighboring samples need to have for a clean re-labelling.
    This evaluation is based on the label-wise precision and recall results of our graph-based re-labelling method as a function of a threshold on the minimum number of nearest neighbors containing the same label.
    Results suggest that precision and recall increase until this threshold is between 4 and 6 nearest neighbours, and plateaus afterwards. Hence, when 4 to 6 neighbours (depending on the noisy rate) share a particular label, it is probable that the noisy training sample has this clean label.

    \subsection{Ablation Study}
    
    \noindent\textbf{Language models}. Our ablation study starts with an investigation of the language models for our BoMD, where we consider three types of models:
    1) a randomly initialised model (without relying on language models); 2) a computer vision language model (e.g., Common Crawl data~\footnote{\url{https://commoncrawl.org}}); and 3) a medical language model (e.g., BioBERT\footnote{Biomedical language model, pretrained on PubMed~\cite{lee2020biobert}.}, ClinicalBERT\footnote{MIMIC corpus (FT on BioBERT)~\cite{huang2019clinicalbert}.}, BlueBERT\footnote{Pretrained on PubMed abstract + MIMIC-III (clinical notes)~\cite{peng2019transfer}.}).
    In~\cref{tab:ablation_pretrain}, the BERTs box shows that BERT models enable better performance than other models, with gains from $2.50\%$ to $1.50\%$ on OpenI and PadChest.
    We argue that this is because the word embeddings from BERT models contain relevant clinical semantic meaning (e.g., Atelectasis and Pneumonia are both correlated to lung opacity, but uncorrelated with Enlarged Cardiomediastinal~\cite{cxp}) that facilitates the multi-label descriptor learning of our method. 
    Among the models trained on BERT models, we observe small variations, which can be related to: 1) the size of the training set, and 2) the relatedness of the medical dataset to our CXR classification problem (BlueBERT is arguably more related than BioBERT).
    
    \noindent\textbf{Evaluation of MID}. In~\cref{tab:ablation_pretrain} the Stage-one training box shows a study of the effectiveness of MID for graph construction and downstream classification task, by comparing it against the use of the descriptors from NVUM~\cite{liu2021nvum}. 
    Given that NVUM produces one descriptor per image, we set MID's number of descriptors per image at $M=1$ for fairness.
    Results from the table show that MID descriptors allow an improvement of between 2\% and 3\% compared to NVUM's descriptors. In addition, $M=1$ represents a graph with one (instead of multiple) descriptor per image, where we can observe a 1\% to 2\% drop, which indicates that our aggregating sub-graph is a more suitable strategy for multi-label images.

    \begin{table}[t]
        \centering

        \resizebox{0.9\columnwidth}{!}{%
        \begin{tabular}{c|c|c|c|c}
        \toprule
        Ablation Study   & $M$ & Language Models & Open-i & PadChest \\\midrule\midrule
        \multirow{5}{*}{BERTs} & 3 & Random Init. & 87.02 & 84.99 \\ 
        ~    & 3 & Glove~\cite{pennington2014glove} & 87.62 & 85.08 \\ 
        ~    & 3 & ClinicalBERT~\cite{huang2019clinicalbert} & 88.27 & 85.72 \\ 
        ~    & 3 & BioBERT~\cite{lee2020biobert} & 89.11 & 86.27 \\ 
        ~    & 3 & BlueBERT~\cite{peng2019transfer} & \textbf{89.52} & \textbf{86.50} \\ \midrule
        \multirow{3}{*}{Stage-one training} 
        & 1 & Self-supervised~\cite{zhou2022conditional} & 84.50 & 83.21 \\
        & 1 & NVUM~\cite{liu2021nvum} & 86.69 & 84.66 \\
        ~ & 1 & MID & 88.34 & 86.02 \\ 
        ~ & 3 & MID & \textbf{89.52} & \textbf{87.50} \\
        \bottomrule
        \end{tabular}
        }
        \vspace{-5pt}
        \caption{Ablation study that compares the mean testing AUC results of our BoMD with the use of different language models (BERTs) and descriptor training (Stage-one training), $M$ shows the number of descriptors per image.}
        \vspace{-10pt}
        \label{tab:ablation_pretrain}
    \end{table}
    
    \begin{table}[t]
        \centering
        \resizebox{0.5\columnwidth}{!}{%
        \begin{tabular}{ c|c|c|c|c}
        \toprule
           $\mathbf{1}$  	    & $\bar{\mathbf{y}}$       	         	&	$\mathbf{m}$	            &	OpenI  &	PadChest  \\\midrule\midrule
            \checkmark 	&	     &		            &	83.72   &	80.93 \\
            \checkmark	&	\checkmark     &		            &	87.92   &	85.48 \\
            	&	\checkmark	           &    \checkmark	    &	89.11	&	86.27 \\
            \checkmark	&   \checkmark     &	\checkmark	    &	\textbf{89.52}	&	\textbf{86.50} \\ \bottomrule
        \end{tabular}
        }
        \vspace{-5pt}
        \caption{Ablation study of the testing AUC results of the  components of our re-labelling in~\eqref{eq:mixup}. $\bar{\mathbf{y}}$ indicates the KNN propagated label, $\mathbf{1}$ is the uniform distribution, and $\mathbf{m}$ is the binary mask.}
        \label{tab:ablation_mixup}
        \vspace{-10pt}
    \end{table}
    
    \noindent\textbf{Smoothly re-labelling}. We show an ablation study of the mixup terms in~\cref{eq:mixup} in terms of the testing AUC results in~\cref{tab:ablation_mixup}. 
    First, we mix up $\mathbf{y}$ and the uniform distribution $\mathbf{1}$ (i.e., label smoothing) with a fixed mixup coefficient of $\lambda=0.6$ (first row of~\cref{tab:ablation_mixup}), then we introduce $\bar{\mathbf{y}}$ with another fixed mixup coefficient of $\gamma=0.25$ (second row of~\cref{tab:ablation_mixup}) and observe improvements of over $4\%$. Next, we remove $\mathbf{1}$ and add the binary mask $\mathbf{m}$ (third row of~\cref{tab:ablation_mixup}) to filter out the confident negative labels, which increases the performance by around $1\%$. The integration of all re-labelling components further increases  performance from $0.23\%$ to $0.41\%$. These results suggest that the mask $\mathbf{m}$ combined with the KNN average label $\bar{\mathbf{y}}$ mitigate the over-smoothing promoted by the uniform distribution $\mathbf{1}$.

    \noindent\textbf{Pre-training with vision-language model~\cite{zhou2022conditional}}. 
    CoOp~\cite{zhou2022conditional} is an effective visual-textual pre-training that can be considered for improving any of the general methods in Tab.~\ref{tab:nih_result} and~\ref{tab:cxp_result}. Hence, we pre-trained with CoOp the method in~\cite{hermoza2020region} and noticed a $1.04\%$ and $0.69\%$ performance drop in Tab.~\ref{tab:nih_result} and~\ref{tab:cxp_result}. 
    This drop can be explained by the fact that CoOP requires backpropagation for the context tokens of the labels during  training, where noisy labels may have caused the learned token to be inaccurate.

    \noindent \textbf{Additional results.}
    In the supplementary material, we include a visualisation of different label smoothing techniques using a t-SNE~\cite{van2008visualizing} graph for a toy problem, additional evaluation for descriptors extracted by the MID module, and detailed sensitivity testing of hyper-parameters.

\section{Discussion and Conclusion}

In this work, we proposed BoMD, a new method to learn from noisy multi-label CXR datasets.
BoMD explores the clinical semantic information, represented by word embeddings from BlueBERT~\cite{peng2019transfer}, to optimise the multi-label image descriptors which are used to find noisy multi-label training samples. We then use the learned image descriptors to build a graph for smoothly re-labelling the training data.
BoMD outperforms current SOTA methods on three real-world CXR benchmarks that consist of training on two large-scale noisy multi-label CXR datasets and testing on three clean multi-label CXR datasets. We additionally evaluate BoMD on our proposed systematic benchmark to further show the effectiveness and robustness of our method.

\textbf{Limitations and future work.} We identify three limitations of BoMD. The first issue is the longer training time (+8h compared with NVUM~\cite{liu2021nvum}) since it requires multiple training stages.  We plan to tackle this problem by better integrating the training stages.
The second issue is that BoMD decreases its performance under extremely noisy label setup (i.e., [0.6, 0.6] in~\cref{tab:noise_auc}), which is due to mistakes in the smooth re-labelling. However, such a high noise rate may not be applicable in real-world scenarios since the usual F1 score for text mining performance is between $80\%$ to $94\%$ ~\cite{nih,openi,oakden2017exploring}, which suggests noise rates much smaller than 60\%.
Another drawback of BoMD is that it does not address imbalanced learning, which is an important point when training with CXR datasets. 
BoMD demonstrates the possibility of leveraging semantic information and sample graphs to estimate the label distribution for training a better CXR classifier. However, noisy-cleaning methods are still dominant in the LNL under multi-class scenarios. One potential future direction is to study a unified framework for addressing both multi-class and multi-label tasks with minimum adaptation.

\section{Acknowledgement}
This work was supported by funding from the Australian Research Council through grant FT190100525.

{\small
\bibliographystyle{ieee_fullname}
\bibliography{egbib}

\begin{thebibliography}{10}\itemsep=-1pt

\bibitem{bae2022noisy}
HeeSun Bae, Seungjae Shin, Byeonghu Na, JoonHo Jang, Kyungwoo Song, and Il-Chul
  Moon.
\newblock From noisy prediction to true label: Noisy prediction calibration via
  generative model.
\newblock In {\em International Conference on Machine Learning}, pages
  1277--1297. PMLR, 2022.

\bibitem{bahri2020deep}
Dara Bahri, Heinrich Jiang, and Maya Gupta.
\newblock Deep k-nn for noisy labels.
\newblock In {\em International Conference on Machine Learning}, pages
  540--550. PMLR, 2020.

\bibitem{ben2021semantic}
Avi Ben-Cohen, Nadav Zamir, Emanuel Ben-Baruch, Itamar Friedman, and Lihi
  Zelnik-Manor.
\newblock Semantic diversity learning for zero-shot multi-label classification.
\newblock In {\em Proceedings of the IEEE/CVF International Conference on
  Computer Vision}, pages 640--650, 2021.

\bibitem{berthelot2019mixmatch}
David Berthelot, Nicholas Carlini, Ian Goodfellow, Nicolas Papernot, Avital
  Oliver, and Colin~A Raffel.
\newblock Mixmatch: A holistic approach to semi-supervised learning.
\newblock {\em Advances in Neural Information Processing Systems}, 32, 2019.

\bibitem{burges2005learning}
Chris Burges, Tal Shaked, Erin Renshaw, Ari Lazier, Matt Deeds, Nicole
  Hamilton, and Greg Hullender.
\newblock Learning to rank using gradient descent.
\newblock In {\em Proceedings of the 22nd international conference on Machine
  learning}, pages 89--96, 2005.

\bibitem{padchest}
Aurelia Bustos, Antonio Pertusa, Jose-Maria Salinas, and Maria de~la
  Iglesia-Vay{\'a}.
\newblock Padchest: A large chest x-ray image dataset with multi-label
  annotated reports.
\newblock {\em Medical image analysis}, 66:101797, 2020.

\bibitem{cohen2021torchxrayvision}
Joseph~Paul Cohen, Joseph~D Viviano, Paul Bertin, Paul Morrison, Parsa
  Torabian, Matteo Guarrera, Matthew~P Lungren, Akshay Chaudhari, Rupert
  Brooks, Mohammad Hashir, et~al.
\newblock Torchxrayvision: A library of chest x-ray datasets and models.
\newblock {\em arXiv preprint arXiv:2111.00595}, 2021.

\bibitem{couairon2022embedding}
Guillaume Couairon, Matthijs Douze, Matthieu Cord, and Holger Schwenk.
\newblock Embedding arithmetic of multimodal queries for image retrieval.
\newblock In {\em Proceedings of the IEEE/CVF Conference on Computer Vision and
  Pattern Recognition}, pages 4950--4958, 2022.

\bibitem{csurka2004visual}
Gabriella Csurka, Christopher Dance, Lixin Fan, Jutta Willamowski, and
  C{\'e}dric Bray.
\newblock Visual categorization with bags of keypoints.
\newblock {\em Workshop on statistical learning in computer vision, ECCV},
  1(1-22):1--2, 2004.

\bibitem{openi}
Dina Demner-Fushman, Marc~D Kohli, Marc~B Rosenman, Sonya~E Shooshan, Laritza
  Rodriguez, Sameer Antani, George~R Thoma, and Clement~J McDonald.
\newblock Preparing a collection of radiology examinations for distribution and
  retrieval.
\newblock {\em Journal of the American Medical Informatics Association},
  23(2):304--310, 2016.

\bibitem{englesson2021generalized}
Erik Englesson and Hossein Azizpour.
\newblock Generalized jensen-shannon divergence loss for learning with noisy
  labels.
\newblock {\em Advances in Neural Information Processing Systems},
  34:30284--30297, 2021.

\bibitem{fiorini2018user}
Nicolas Fiorini, Robert Leaman, David~J Lipman, and Zhiyong Lu.
\newblock How user intelligence is improving pubmed.
\newblock {\em Nature biotechnology}, 36(10):937--945, 2018.

\bibitem{goldberger2016training}
Jacob Goldberger and Ehud Ben-Reuven.
\newblock Training deep neural-networks using a noise adaptation layer.
\newblock {\em 4th International Conference on Learning Representations
  (ICLR)}, 2016.

\bibitem{han2018co}
Bo Han, Quanming Yao, Xingrui Yu, Gang Niu, Miao Xu, Weihua Hu, Ivor Tsang, and
  Masashi Sugiyama.
\newblock Co-teaching: Robust training of deep neural networks with extremely
  noisy labels.
\newblock {\em Advances in neural information processing systems}, 31, 2018.

\bibitem{harris1954distributional}
Zellig~S Harris.
\newblock Distributional structure.
\newblock {\em Word}, 10(2-3):146--162, 1954.

\bibitem{hermoza2020region}
Renato Hermoza, Gabriel Maicas, Jacinto~C Nascimento, and Gustavo Carneiro.
\newblock Region proposals for saliency map refinement for weakly-supervised
  disease localisation and classification.
\newblock In {\em International Conference on Medical Image Computing and
  Computer-Assisted Intervention}, pages 539--549. Springer, 2020.

\bibitem{hu2018learning}
Hexiang Hu, Wei-Lun Chao, and Fei Sha.
\newblock Learning answer embeddings for visual question answering.
\newblock In {\em Proceedings of the IEEE Conference on Computer Vision and
  Pattern Recognition}, pages 5428--5436, 2018.

\bibitem{huang2017densely}
Gao Huang, Zhuang Liu, Laurens Van Der~Maaten, and Kilian~Q Weinberger.
\newblock Densely connected convolutional networks.
\newblock In {\em Proceedings of the IEEE conference on computer vision and
  pattern recognition}, pages 4700--4708, 2017.

\bibitem{huang2019o2u}
Jinchi Huang, Lie Qu, Rongfei Jia, and Binqiang Zhao.
\newblock O2u-net: A simple noisy label detection approach for deep neural
  networks.
\newblock In {\em Proceedings of the IEEE/CVF International Conference on
  Computer Vision}, pages 3326--3334, 2019.

\bibitem{huang2019clinicalbert}
Kexin Huang, Jaan Altosaar, and Rajesh Ranganath.
\newblock Clinicalbert: Modeling clinical notes and predicting hospital
  readmission.
\newblock {\em arXiv preprint arXiv:1904.05342}, 2019.

\bibitem{cxp}
Jeremy Irvin et~al.
\newblock Chexpert: A large chest radiograph dataset with uncertainty labels
  and expert comparison.
\newblock In {\em AAAI}, volume~33, pages 590--597, 2019.

\bibitem{iscen2022learning}
Ahmet Iscen, Jack Valmadre, Anurag Arnab, and Cordelia Schmid.
\newblock Learning with neighbor consistency for noisy labels.
\newblock In {\em Proceedings of the IEEE/CVF Conference on Computer Vision and
  Pattern Recognition}, pages 4672--4681, 2022.

\bibitem{jiang2018mentornet}
Lu Jiang, Zhengyuan Zhou, Thomas Leung, Li-Jia Li, and Li Fei-Fei.
\newblock Mentornet: Learning data-driven curriculum for very deep neural
  networks on corrupted labels.
\newblock In {\em ICML}, 2018.

\bibitem{johnson2016mimic}
Alistair~EW Johnson, Tom~J Pollard, Lu Shen, Li-wei~H Lehman, Mengling Feng,
  Mohammad Ghassemi, Benjamin Moody, Peter Szolovits, Leo Anthony~Celi, and
  Roger~G Mark.
\newblock Mimic-iii, a freely accessible critical care database.
\newblock {\em Scientific data}, 3(1):1--9, 2016.

\bibitem{johnson2019billion}
Jeff Johnson, Matthijs Douze, and Herv{\'e} J{\'e}gou.
\newblock Billion-scale similarity search with {GPUs}.
\newblock {\em IEEE Transactions on Big Data}, 7(3):535--547, 2019.

\bibitem{kim2021fine}
Taehyeon Kim, Jongwoo Ko, JinHwan Choi, Se-Young Yun, et~al.
\newblock Fine samples for learning with noisy labels.
\newblock {\em Advances in Neural Information Processing Systems},
  34:24137--24149, 2021.

\bibitem{kingma2014adam}
Diederik~P Kingma and Jimmy Ba.
\newblock Adam: A method for stochastic optimization.
\newblock {\em arXiv preprint arXiv:1412.6980}, 2014.

\bibitem{lee2020biobert}
Jinhyuk Lee, Wonjin Yoon, Sungdong Kim, Donghyeon Kim, Sunkyu Kim, Chan~Ho So,
  and Jaewoo Kang.
\newblock Biobert: a pre-trained biomedical language representation model for
  biomedical text mining.
\newblock {\em Bioinformatics}, 36(4):1234--1240, 2020.

\bibitem{li2020dividemix}
Junnan Li, Richard Socher, and Steven~CH Hoi.
\newblock Dividemix: Learning with noisy labels as semi-supervised learning.
\newblock {\em arXiv preprint arXiv:2002.07394}, 2020.

\bibitem{li2022estimating}
Shikun Li, Xiaobo Xia, Hansong Zhang, Yibing Zhan, Shiming Ge, and Tongliang
  Liu.
\newblock Estimating noise transition matrix with label correlations for noisy
  multi-label learning.
\newblock {\em Advances in Neural Information Processing Systems},
  35:24184--24198, 2022.

\bibitem{litjens2017survey}
Geert Litjens, Thijs Kooi, Babak~Ehteshami Bejnordi, Arnaud Arindra~Adiyoso
  Setio, Francesco Ciompi, Mohsen Ghafoorian, Jeroen~Awm Van Der~Laak, Bram
  Van~Ginneken, and Clara~I S{\'a}nchez.
\newblock A survey on deep learning in medical image analysis.
\newblock {\em Medical image analysis}, 42:60--88, 2017.

\bibitem{liu2021nvum}
Fengbei Liu, Yuanhong Chen, Yu Tian, Yuyuan Liu, Chong Wang, Vasileios
  Belagiannis, and Gustavo Carneiro.
\newblock Nvum: Non-volatile unbiased memory for robust medical image
  classification.
\newblock {\em arXiv e-prints}, pages arXiv--2103, 2021.

\bibitem{liu2021acpl}
Fengbei Liu, Yu Tian, Yuanhong Chen, Yuyuan Liu, Vasileios Belagiannis, and
  Gustavo Carneiro.
\newblock Acpl: Anti-curriculum pseudo-labelling for semi-supervised medical
  image classification.
\newblock {\em arXiv preprint arXiv:2111.12918}, 2021.

\bibitem{liu2020early}
Sheng Liu, Jonathan Niles-Weed, Narges Razavian, and Carlos Fernandez-Granda.
\newblock Early-learning regularization prevents memorization of noisy labels.
\newblock {\em Advances in neural information processing systems},
  33:20331--20342, 2020.

\bibitem{liu2022perturbed}
Yuyuan Liu, Yu Tian, Yuanhong Chen, Fengbei Liu, Vasileios Belagiannis, and
  Gustavo Carneiro.
\newblock Perturbed and strict mean teachers for semi-supervised semantic
  segmentation.
\newblock In {\em Proceedings of the IEEE/CVF Conference on Computer Vision and
  Pattern Recognition}, pages 4258--4267, 2022.

\bibitem{liu2022translation}
Yuyuan Liu, Yu Tian, Chong Wang, Yuanhong Chen, Fengbei Liu, Vasileios
  Belagiannis, and Gustavo Carneiro.
\newblock Translation consistent semi-supervised segmentation for 3d medical
  images.
\newblock {\em arXiv preprint arXiv:2203.14523}, 2022.

\bibitem{loshchilov2016sgdr}
Ilya Loshchilov and Frank Hutter.
\newblock Sgdr: Stochastic gradient descent with warm restarts.
\newblock {\em arXiv preprint arXiv:1608.03983}, 2016.

\bibitem{lowe1999object}
David~G Lowe.
\newblock Object recognition from local scale-invariant features.
\newblock In {\em Proceedings of the seventh IEEE international conference on
  computer vision}, volume~2, pages 1150--1157. Ieee, 1999.

\bibitem{lukasik2020does}
Michal Lukasik, Srinadh Bhojanapalli, Aditya Menon, and Sanjiv Kumar.
\newblock Does label smoothing mitigate label noise?
\newblock In {\em International Conference on Machine Learning}, pages
  6448--6458. PMLR, 2020.

\bibitem{ma2019multi}
Congbo Ma, Hu Wang, and Steven~CH Hoi.
\newblock Multi-label thoracic disease image classification with
  cross-attention networks.
\newblock In {\em International Conference on Medical Image Computing and
  Computer-Assisted Intervention}, pages 730--738. Springer, 2019.

\bibitem{ma2020normalized}
Xingjun Ma, Hanxun Huang, Yisen Wang, Simone Romano, Sarah Erfani, and James
  Bailey.
\newblock Normalized loss functions for deep learning with noisy labels.
\newblock In {\em International conference on machine learning}, pages
  6543--6553. PMLR, 2020.

\bibitem{majkowska2020chest}
Anna Majkowska, Sid Mittal, David~F Steiner, Joshua~J Reicher, Scott~Mayer
  McKinney, Gavin~E Duggan, Krish Eswaran, Po-Hsuan Cameron~Chen, Yun Liu,
  Sreenivasa~Raju Kalidindi, et~al.
\newblock Chest radiograph interpretation with deep learning models: assessment
  with radiologist-adjudicated reference standards and population-adjusted
  evaluation.
\newblock {\em Radiology}, 294(2):421--431, 2020.

\bibitem{malach2017decoupling}
Eran Malach and Shai Shalev-Shwartz.
\newblock Decoupling" when to update" from" how to update".
\newblock {\em Advances in Neural Information Processing Systems}, 30, 2017.

\bibitem{muller2019does}
Rafael M{\"u}ller, Simon Kornblith, and Geoffrey~E Hinton.
\newblock When does label smoothing help?
\newblock {\em Advances in neural information processing systems}, 32, 2019.

\bibitem{oakden2017exploring}
Luke Oakden-Rayner.
\newblock Exploring the chestxray14 dataset: problems.
\newblock {\em Wordpress: Luke Oakden Rayner}, 2017.

\bibitem{oakden2020exploring}
Luke Oakden-Rayner.
\newblock Exploring large-scale public medical image datasets.
\newblock {\em Academic radiology}, 27(1):106--112, 2020.

\bibitem{paszke2019pytorch}
Adam Paszke, Sam Gross, Francisco Massa, Adam Lerer, James Bradbury, Gregory
  Chanan, Trevor Killeen, Zeming Lin, Natalia Gimelshein, Luca Antiga, et~al.
\newblock Pytorch: An imperative style, high-performance deep learning library.
\newblock {\em Advances in neural information processing systems}, 32, 2019.

\bibitem{patrini2017making}
Giorgio Patrini, Alessandro Rozza, Aditya Krishna~Menon, Richard Nock, and
  Lizhen Qu.
\newblock Making deep neural networks robust to label noise: A loss correction
  approach.
\newblock In {\em Proceedings of the IEEE conference on computer vision and
  pattern recognition}, pages 1944--1952, 2017.

\bibitem{peng2019transfer}
Yifan Peng, Shankai Yan, and Zhiyong Lu.
\newblock Transfer learning in biomedical natural language processing: an
  evaluation of bert and elmo on ten benchmarking datasets.
\newblock {\em arXiv preprint arXiv:1906.05474}, 2019.

\bibitem{pennington2014glove}
Jeffrey Pennington, Richard Socher, and Christopher~D Manning.
\newblock Glove: Global vectors for word representation.
\newblock In {\em Proceedings of the 2014 conference on empirical methods in
  natural language processing (EMNLP)}, pages 1532--1543, 2014.

\bibitem{rajpurkar2017chexnet}
Pranav Rajpurkar, Jeremy Irvin, Kaylie Zhu, Brandon Yang, Hershel Mehta, Tony
  Duan, Daisy Ding, Aarti Bagul, Curtis Langlotz, Katie Shpanskaya, et~al.
\newblock Chexnet: Radiologist-level pneumonia detection on chest x-rays with
  deep learning.
\newblock {\em arXiv preprint arXiv:1711.05225}, 2017.

\bibitem{ridnik2021asymmetric}
Tal Ridnik, Emanuel Ben-Baruch, Nadav Zamir, Asaf Noy, Itamar Friedman, Matan
  Protter, and Lihi Zelnik-Manor.
\newblock Asymmetric loss for multi-label classification.
\newblock In {\em Proceedings of the IEEE/CVF International Conference on
  Computer Vision}, pages 82--91, 2021.

\bibitem{russakovsky2015imagenet}
Olga Russakovsky, Jia Deng, Hao Su, Jonathan Krause, Sanjeev Satheesh, Sean Ma,
  Zhiheng Huang, Andrej Karpathy, Aditya Khosla, Michael Bernstein, et~al.
\newblock Imagenet large scale visual recognition challenge.
\newblock {\em International journal of computer vision}, 115(3):211--252,
  2015.

\bibitem{sivic2003video}
Josef Sivic and Andrew Zisserman.
\newblock Video google: A text retrieval approach to object matching in videos.
\newblock In {\em Computer Vision, IEEE International Conference on}, volume~3,
  pages 1470--1470. IEEE Computer Society, 2003.

\bibitem{sivic2008efficient}
Josef Sivic and Andrew Zisserman.
\newblock Efficient visual search of videos cast as text retrieval.
\newblock {\em IEEE transactions on pattern analysis and machine intelligence},
  31(4):591--606, 2008.

\bibitem{szegedy2016rethinking}
Christian Szegedy, Vincent Vanhoucke, Sergey Ioffe, Jon Shlens, and Zbigniew
  Wojna.
\newblock Rethinking the inception architecture for computer vision.
\newblock In {\em Proceedings of the IEEE conference on computer vision and
  pattern recognition}, pages 2818--2826, 2016.

\bibitem{van2008visualizing}
Laurens Van~der Maaten and Geoffrey Hinton.
\newblock Visualizing data using t-sne.
\newblock {\em Journal of machine learning research}, 9(11), 2008.

\bibitem{nih}
Xiaosong Wang, Yifan Peng, Le Lu, Zhiyong Lu, Mohammadhadi Bagheri, and
  Ronald~M Summers.
\newblock Chestx-ray8: Hospital-scale chest x-ray database and benchmarks on
  weakly-supervised classification and localization of common thorax diseases.
\newblock In {\em Proceedings of the IEEE conference on computer vision and
  pattern recognition}, pages 2097--2106, 2017.

\bibitem{wang2019symmetric}
Yisen Wang, Xingjun Ma, Zaiyi Chen, Yuan Luo, Jinfeng Yi, and James Bailey.
\newblock Symmetric cross entropy for robust learning with noisy labels.
\newblock In {\em Proceedings of the IEEE/CVF International Conference on
  Computer Vision}, pages 322--330, 2019.

\bibitem{wei2021smooth}
Jiaheng Wei, Hangyu Liu, Tongliang Liu, Gang Niu, Masashi Sugiyama, and Yang
  Liu.
\newblock To smooth or not? when label smoothing meets noisy labels.
\newblock {\em Learning}, 1(1):e1, 2021.

\bibitem{wu2020topological}
Pengxiang Wu, Songzhu Zheng, Mayank Goswami, Dimitris Metaxas, and Chao Chen.
\newblock A topological filter for learning with label noise.
\newblock {\em Advances in neural information processing systems},
  33:21382--21393, 2020.

\bibitem{wu2021ngc}
Zhi-Fan Wu, Tong Wei, Jianwen Jiang, Chaojie Mao, Mingqian Tang, and Yu-Feng
  Li.
\newblock Ngc: a unified framework for learning with open-world noisy data.
\newblock In {\em Proceedings of the IEEE/CVF International Conference on
  Computer Vision}, pages 62--71, 2021.

\bibitem{xia2020part}
Xiaobo Xia, Tongliang Liu, Bo Han, Nannan Wang, Mingming Gong, Haifeng Liu,
  Gang Niu, Dacheng Tao, and Masashi Sugiyama.
\newblock Part-dependent label noise: Towards instance-dependent label noise.
\newblock {\em Advances in Neural Information Processing Systems},
  33:7597--7610, 2020.

\bibitem{xie2021partial}
Ming-Kun Xie and Sheng-Jun Huang.
\newblock Partial multi-label learning with noisy label identification.
\newblock {\em IEEE Transactions on Pattern Analysis and Machine Intelligence},
  44(7):3676--3687, 2021.

\bibitem{xue2022robust}
Cheng Xue, Lequan Yu, Pengfei Chen, Qi Dou, and Pheng-Ann Heng.
\newblock Robust medical image classification from noisy labeled data with
  global and local representation guided co-training.
\newblock {\em IEEE Transactions on Medical Imaging}, 41(6):1371--1382, 2022.

\bibitem{yao2020dual}
Yu Yao, Tongliang Liu, Bo Han, Mingming Gong, Jiankang Deng, Gang Niu, and
  Masashi Sugiyama.
\newblock Dual t: Reducing estimation error for transition matrix in
  label-noise learning.
\newblock {\em Advances in neural information processing systems},
  33:7260--7271, 2020.

\bibitem{zhang2021delving}
Chang-Bin Zhang, Peng-Tao Jiang, Qibin Hou, Yunchao Wei, Qi Han, Zhen Li, and
  Ming-Ming Cheng.
\newblock Delving deep into label smoothing.
\newblock {\em IEEE Transactions on Image Processing}, 30:5984--5996, 2021.

\bibitem{zhang2016fast}
Yang Zhang, Boqing Gong, and Mubarak Shah.
\newblock Fast zero-shot image tagging.
\newblock In {\em 2016 IEEE Conference on Computer Vision and Pattern
  Recognition (CVPR)}, pages 5985--5994. IEEE, 2016.

\bibitem{zhang2018generalized}
Zhilu Zhang and Mert Sabuncu.
\newblock Generalized cross entropy loss for training deep neural networks with
  noisy labels.
\newblock {\em Advances in neural information processing systems}, 31, 2018.

\bibitem{zhao2021evaluating}
Wenting Zhao and Carla Gomes.
\newblock Evaluating multi-label classifiers with noisy labels.
\newblock {\em arXiv preprint arXiv:2102.08427}, 2021.

\bibitem{zhou2022conditional}
Kaiyang Zhou, Jingkang Yang, Chen~Change Loy, and Ziwei Liu.
\newblock Conditional prompt learning for vision-language models.
\newblock In {\em Proceedings of the IEEE/CVF Conference on Computer Vision and
  Pattern Recognition}, pages 16816--16825, 2022.

\end{thebibliography}
}

\clearpage
\section{Appendix}

\begin{table*}[t]
    \centering
    \begin{tabularx}{\linewidth}{ l *{4}{|Y} }
    \toprule
                 & \multicolumn{2}{c|}{Train} & \multicolumn{2}{c}{Test} \\ \midrule \midrule
    Datasets     & NIH~\cite{nih}         & CXP~\cite{cxp}         & OpenI~\cite{openi}      & PadChest~\cite{padchest}    \\ \midrule
    Train on NIH & 83,672 (14)  & -           & 2,971 (14)  & 14,714 (14)  \\
    Train on CXP & -           & 170,958 (8)  & 2,823 (8)   & 12,885 (8)  \\
    \bottomrule
    \end{tabularx}
    \captionof{table}{Statistics for all datasets after data pre-processing, where the digit on the left is the total number of samples and the digit inside brackets is the number of classes.}
    \label{tab:dataset}
\end{table*}

\begin{table*}[t]
    \centering
    \resizebox{1\textwidth}{!}{  
    \begin{tabular}{c||c||cc||c||cc||c||cc||c||cc}
    \toprule
    \multirow{1}{*}{Experiments} & \multicolumn{6}{c||}{Mixup Coefficient}                   & \multicolumn{3}{c||}{Number of Descriptors} & \multicolumn{3}{c}{K-nearest neighbour}  \\ \midrule \midrule
    \multirow{1}{*}{Settings}    & $\lambda$    & OpenI             & PadChest & $\gamma$    & OpenI             & PadChest & $M$       & OpenI          & PadChest       & $K$       & OpenI       & PadChest      \\ \midrule
    \multirow{6}{*}{AUC}         
    &	0.2	&	88.39	&	85.52	&	0.05	&	89.14	&	86.05	&	1	&	88.34	&	86.02	&	5	&	89.20	&	86.15	\\
    &	0.4	&	88.56	&	85.93	&	0.15	&	87.87	&	86.17	&	3	&	\textbf{89.52}	&	\textbf{86.50}	&	10	&	\textbf{89.52}	&	\textbf{86.50}	\\
    &	0.6	&	\textbf{89.52}	&	\textbf{86.50}	&	0.25	&	\textbf{89.52}	&	\textbf{86.50}	&	5	&	88.92	&	86.39	&	20	&	88.23	&	85.79	\\
    &	0.8	&	88.37	&	86.29	&	0.35	&	88.40	&	86.48	&	7	&	89.03	&	86.43	&	50	&	87.59	&	85.49	\\
    &	1.0	&	88.31	&	86.21	&	0.45	&	88.46	&	86.46	&	9	&	88.45	&	86.29	&	100	&	87.36	&	85.48	\\
	\bottomrule													
    \end{tabular}
    }
    \captionof{table}{Ablation study of the hyper-parameters using mean AUC. Models are trained on NIH~\cite{nih} and tested on OpenI~\cite{openi} and PadChest~\cite{padchest}. Note that for each hyper-parameter, we fix the others to their best values (i.e., $\lambda=0.6$, $\gamma=0.25$, $M=3$ and $K=10$).}
    \label{tab:ablation}
    \vspace{-5pt}
\end{table*}




\begin{figure*}[t]
    \centering        
    \subfloat[\centering Recall]{{\includegraphics[width=0.35\linewidth]{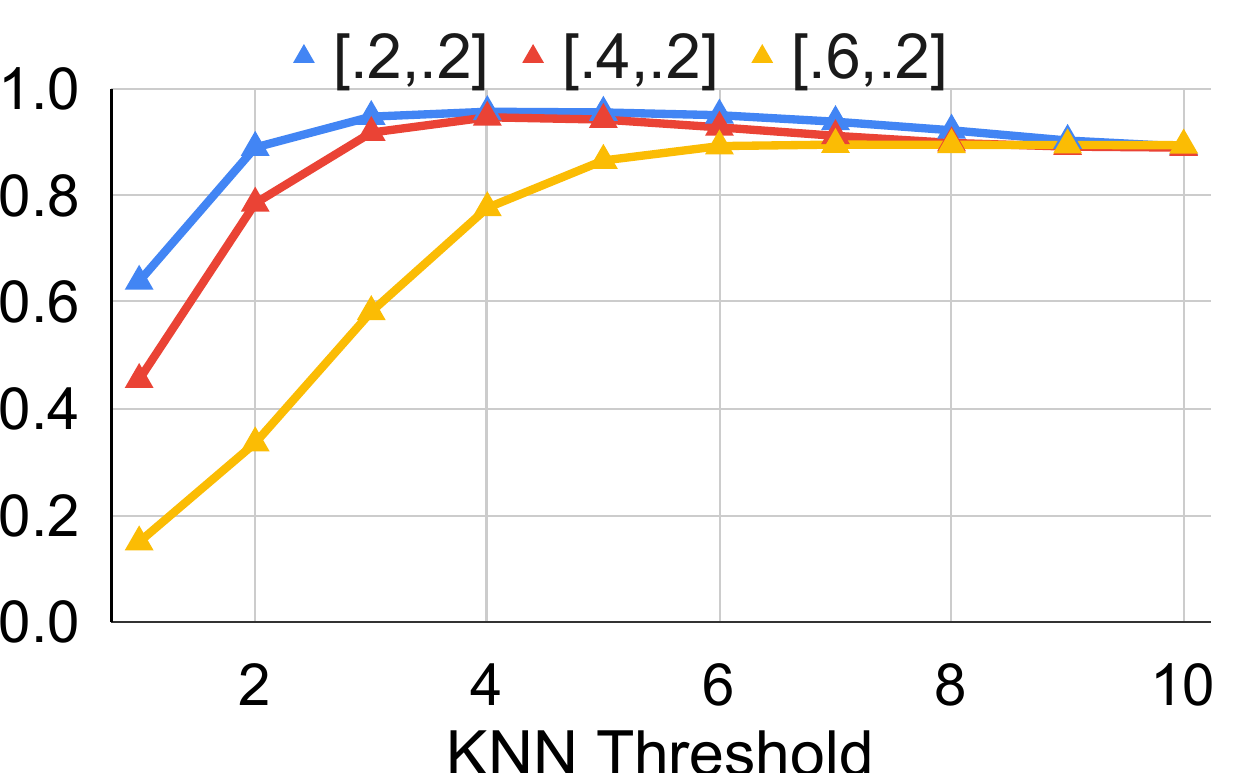} }}\hspace{20pt}
    \subfloat[\centering Precision]{{\includegraphics[width=0.35\linewidth]{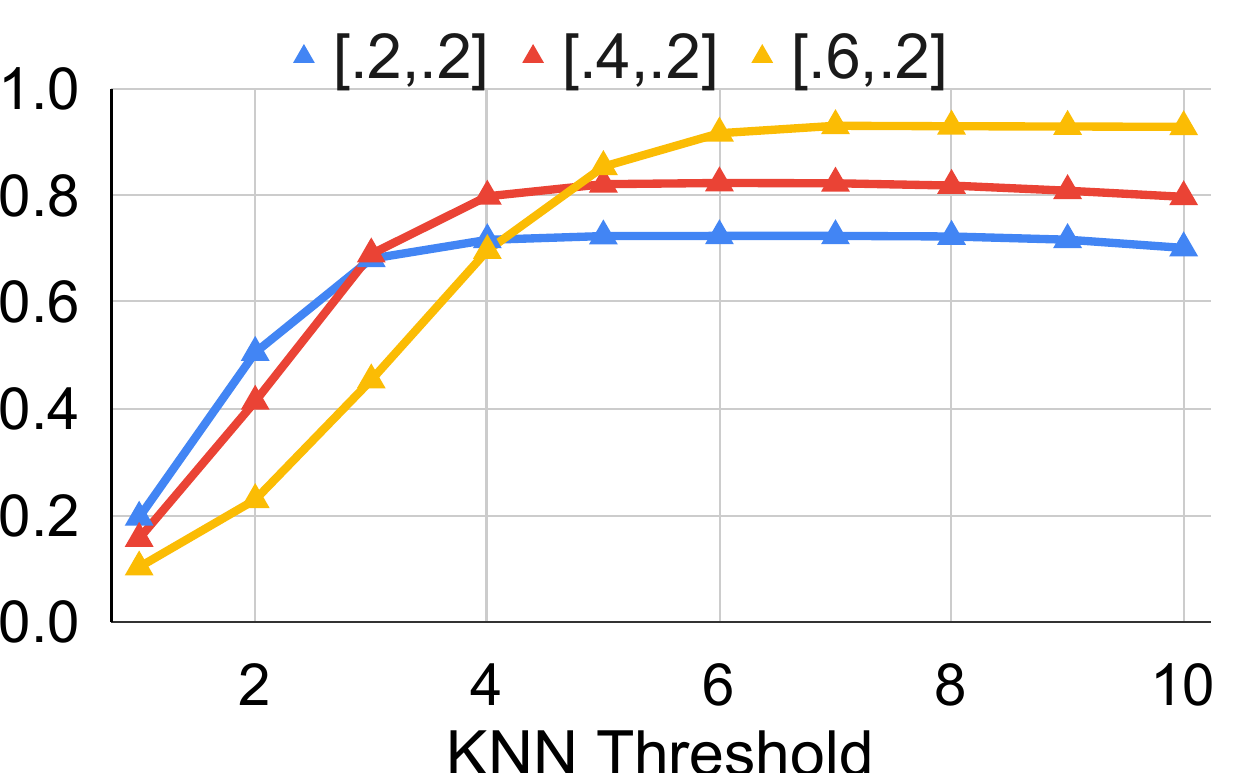} }}%
    \caption{Label-wise precision and recall of our KNN propagated label under $\bar{\mathbf{y}}$ w.r.t the clean annotation from PadChest. The horizontal axis shows a threshold of the minimum number of nearest neighbors containing each class.
    }
    \label{fig:knn_acc}
\end{figure*}

\section{Dataset Statistics}

Table~\ref{tab:dataset} shows the statistics of our training noisy training set (NIH~\cite{nih} and ChestXpert (CXP)~\cite{cxp}) and clean testing sets (OpenI~\cite{openi} and PadChest~\cite{padchest}). 
Due to  inconsistencies in the number of labels for each dataset, we trim the original datasets and only keep the samples that contain labels present in all datasets based on~\cite{cohen2021torchxrayvision,liu2021nvum}.
After our data pre-processing, there are 83,672  frontal-view images with 14 common chest radiographic observations for NIH~\cite{nih} dataset where the corresponding testing sets for OpenI~\cite{openi} and PadChest~\cite{padchest} contain 2,917 and 14,714 frontal-view images respectively. For CXP, we have 170,958  frontal-view images with 8 chest radiographic observations where the corresponding testing set for OpenI~\cite{openi} and PadChest~\cite{padchest} contain 2,823 and 12,885  frontal-view images, respectively.



\begin{figure*}[t!]
    \centering
    \subfloat[\centering Clean \label{fig:clean} ]{{\includegraphics[height=71.5pt]{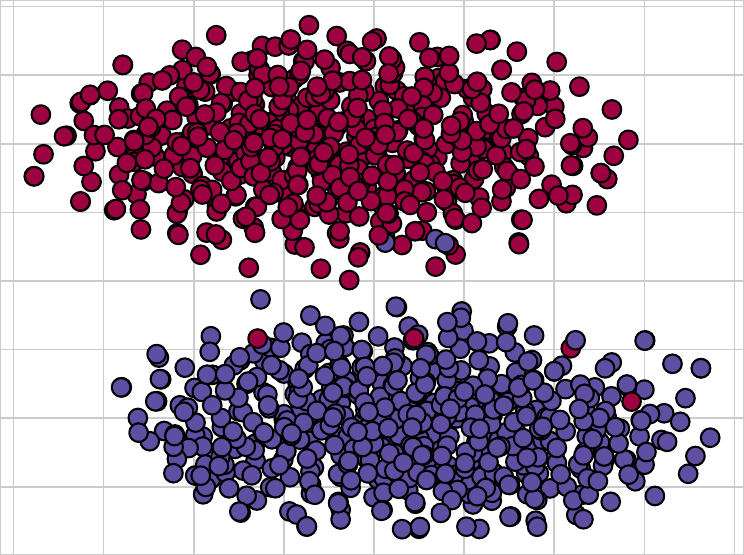} }}%
    \subfloat[\centering Noisy \label{fig:noise}
    ]{{\includegraphics[height=71.5pt]{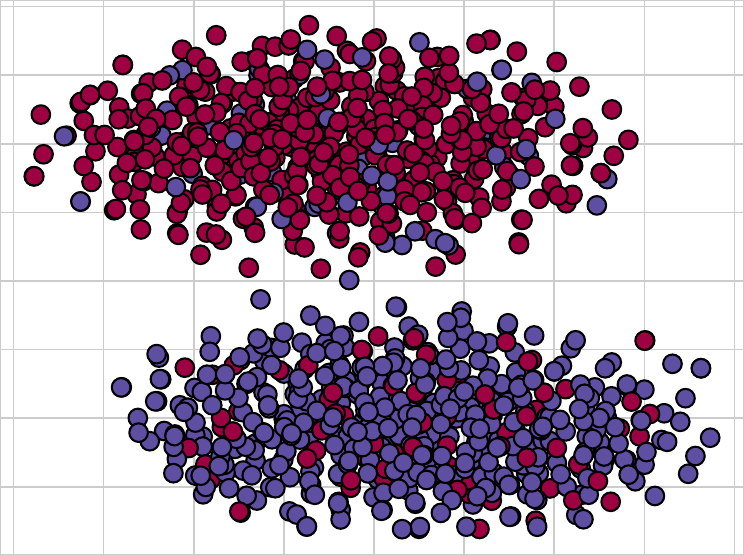} }}%
    \subfloat[\centering LS~\cite{lukasik2020does} \label{fig:ls}
    ]{{\includegraphics[height=71.5pt]{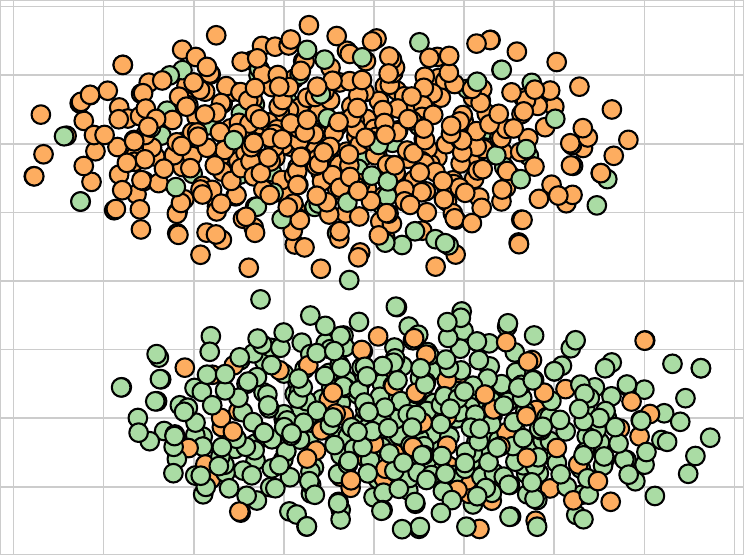} }}%
    \subfloat[\centering BoMD \label{fig:ours} ]{{\includegraphics[height=73.45pt]{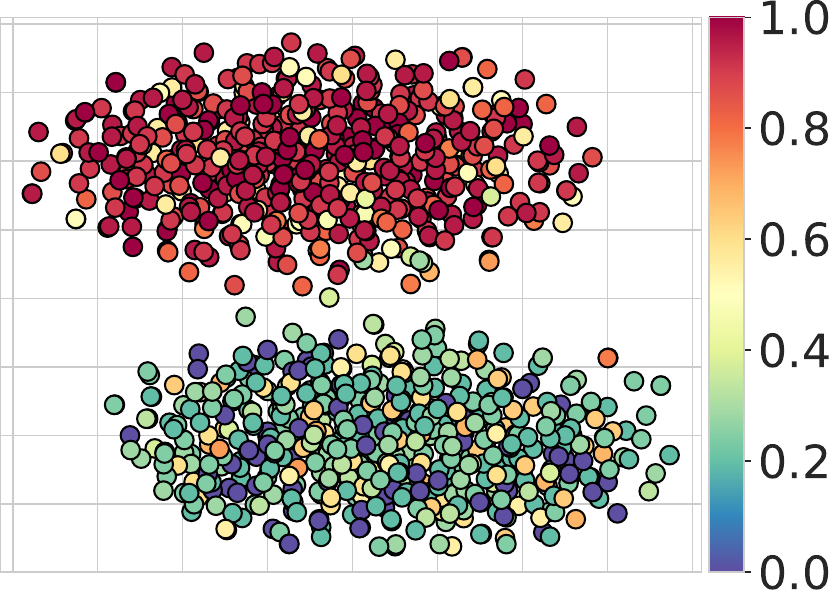}}}%
    \subfloat[\centering GLS~\cite{wei2021smooth} \label{fig:gls} ]{{\includegraphics[height=71.5pt]{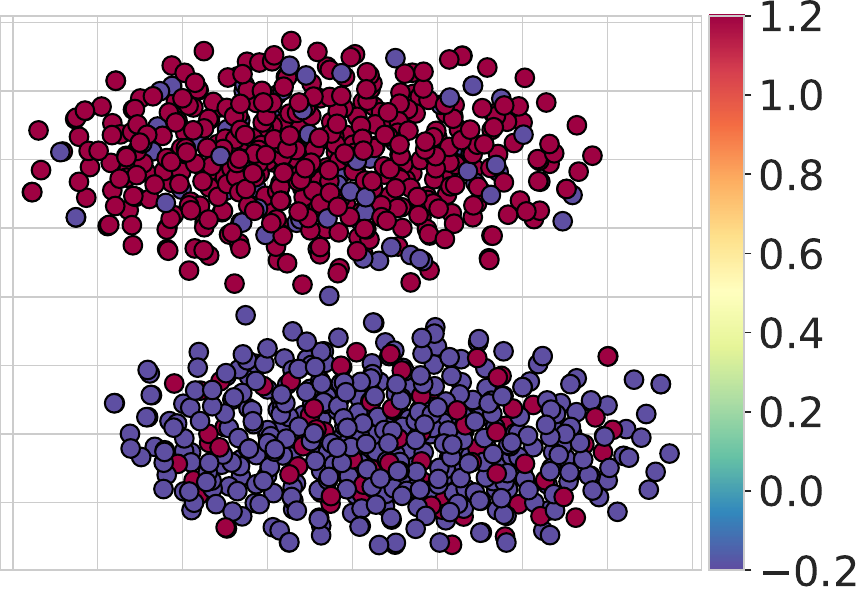} }}%
    \vspace{-5pt}
    \caption{Visualisation of different label smoothing techniques. The color of each data-point indicates the confidence score. We start with two isotropic Gaussian clusters in \textbf{(a)} as the clean set where red points indicate class 1 and blue points represent class 2. We randomly inject $20\%$ of symmetric noise to form the noisy set in \textbf{(b)}. We compare our method (in \textbf{(d)}) with two baseline methods, namely: label smoothing (LS)~\cite{lukasik2020does} (in \textbf{(c)}) and generalised label smoothing (GLS)~\cite{wei2021smooth} (in \textbf{(e)}). We show that our method alleviates the noisy label problem by modifying the confidence score based on the nearest neighbors, while LS pushes the labels toward the uniform distribution and GLS pushes the labels toward sharp binary distribution. Note that GLS has a different scale for confidence scale which is from -0.2 to +1.2, while the others have a range from 0 to 1.
    }%
    \label{fig:tsne}%
    \vspace{-10pt}
\end{figure*}

\section{Further Ablation Studies}

    We evaluate the number of KNN neighboring samples that are required for a clean re-labelling. We measure the precision and recall for the detection of noisy-labels of our graph-based relabelling method in Fig.~\ref{fig:knn_acc} as a function of the threshold of the minimum number of nearest neighbors containing each class.
    For example, if the KNN threshold is 4, then a particular label of a sample is set to 1 only if there are at least 4 neighbors that share the same label.
    Note that the measures are computed in a label-wise manner, and we consider the flipping rate $p_l$ at $20\%$ and the percentage of noisy samples $p_s \in \{20\%, 40\%, 60\%\}$. 
    We observe lower recall rate for lower values of $K$ because the KNN label propagation under multi-label scenario tends to be noisier for small values of $K$. 
    We achieve the highest recall rate when this threshold is between 4 and 6 nearest neighbours, which means that when we have at least 4 samples in the K nearest neighbour that share the same label, it is most likely a true label.

\section{Visualisation of Smoothing Techniques} \label{sec:visual_ls}

    To visualise the performance of different label smoothing techniques, we plot the t-SNE~\cite{van2008visualizing} for a toy problem. 
    More specifically, we first generate two isotropic Gaussian clusters as the clean set (\cref{fig:clean}) and randomly inject $20\%$ of symmetric noise (\cref{fig:noise}) to form a noisy set. We show that our BoMD demonstrates a better tradeoff when correcting the labels since it re-labels the noisy samples without being overconfident in the detection (like shown by GLS~\cite{wei2021smooth}) and without oversmoothing the labels (like displayed by LS~\cite{lukasik2020does}). Note that we set the smoothing parameter $r$ to 0.6 and -0.4 respectively for LS~\cite{lukasik2020does} and GLS~\cite{wei2021smooth}.

\begin{table*}[ht]
    \centering
    \caption{Disease-level testing AUC results for models \textbf{trained on NIH}. 
    }
    \vspace{-5pt}
    \label{tab:nih_result_1}
    \resizebox{1\textwidth}{!}{  
    \begin{tabular}{l|cc|cc|cc|cc|cc|cc}
    \toprule
    Models	
    &	\multicolumn{2}{c|}{Hermoza et al}	&	\multicolumn{2}{c|}{CAN}	&	\multicolumn{2}{c|}{DivideMix}	    &	\multicolumn{2}{c|}{FINE}	
    &	\multicolumn{2}{c|}{ELR}	        &	\multicolumn{2}{c}{NVUM}	
    \\	\midrule	\midrule			
    Datasets	
    &	OpenI	&	PadChest	&	OpenI	&	PadChest	
    &	OpenI	&	PadChest	&	OpenI	&	PadChest	
    &	OpenI	&	PadChest     &	OpenI	&	PadChest 
    \\	\midrule
        Atelectasis & 86.85 & 83.59 & 84.83 & 79.88 & 70.98 & 73.48 & 77.51 & 67.70 & 86.21 & 85.69 & 88.16 & 85.66 \\ 
        Cardiomegaly & 89.49 & 91.25 & 90.87 & 91.72 & 74.74 & 81.63 & 77.93 & 84.54 & 90.79 & 92.81 & 90.57 & 92.94 \\ 
        Effusion & 94.05 & 96.27 & 94.37 & 96.29 & 84.49 & 97.75 & 74.39 & 86.76 & 94.74 & 96.67 & 93.64 & 96.56 \\ 
        Infiltration & 77.48 & 70.61 & 77.88 & 73.78 & 84.03 & 81.61 & 73.41 & 67.28 & 78.92 & 73.82 & 74.30 & 72.51 \\ 
        Mass & 95.72 & 86.93 & 87.47 & 85.81 & 71.31 & 74.41 & 57.45 & 69.54 & 81.90 & 84.51 & 93.06 & 85.93 \\ 
        Nodule & 81.68 & 75.99 & 80.71 & 74.14 & 57.35 & 63.89 & 59.43 & 57.66 & 86.22 & 75.59 & 88.79 & 75.56 \\ 
        Pneumonia & 87.15 & 75.73 & 84.79 & 76.49 & 71.65 & 72.32 & 56.22 & 60.46 & 88.99 & 80.28 & 90.90 & 82.22 \\ 
        Pneumothorax & 75.34 & 74.55 & 82.21 & 79.73 & 75.56 & 75.46 & 59.88 & 64.46 & 78.65 & 78.47 & 85.78 & 79.50 \\ 
        Edema & 84.31 & 97.78 & 82.80 & 96.41 & 80.71 & 85.81 & 58.18 & 95.20 & 85.57 & 97.58 & 86.56 & 95.70 \\ 
        Emphysema & 83.26 & 79.81 & 81.26 & 78.06 & 64.81 & 59.91 & 43.31 & 50.72 & 82.79 & 79.87 & 83.70 & 79.38 \\ 
        Fibrosis & 85.85 & 96.46 & 83.17 & 93.20 & 76.96 & 84.71 & 61.97 & 88.68 & 92.07 & 97.42 & 91.67 & 97.61 \\ 
        Pleural Thicken & 77.99 & 71.85 & 77.59 & 67.87 & 62.98 & 58.25 & 63.17 & 54.33 & 83.45 & 72.01 & 84.82 & 74.80 \\ 
        Hernia & 92.90 & 89.90 & 87.37 & 86.87 & 70.34 & 72.11 & 64.86 & 74.56 & 95.77 & 93.37 & 94.28 & 93.02 \\ 
        Mean AUC & 85.54 & 83.90 & 84.26 & 83.10 & 72.76 & 75.49 & 63.67 & 70.91 & 86.62 & 85.24 & 88.17 & 85.49 \\ 
        \bottomrule
    \end{tabular}
    }
    \vspace{-10pt}
\end{table*}

\begin{table*}[ht]
    \centering
    \caption{Disease-level testing AUC results for models \textbf{trained on NIH}. 
    }
    \vspace{-5pt}
    \label{tab:nih_result_2}
    \resizebox{1\textwidth}{!}{  
    \begin{tabular}{l|cc|cc|cc|cc|cc|cc}
    \toprule
    Models	
    &	\multicolumn{2}{c|}{NPC}	        &	\multicolumn{2}{c|}{NCR}	
    &	\multicolumn{2}{c|}{LS}	            &	\multicolumn{2}{c|}{OLS}	
    &	\multicolumn{2}{c|}{GLS}	        &	\multicolumn{2}{c}{\textbf{BoMD}}
    \\	\midrule	\midrule			
    Datasets	
    &	OpenI	&	PadChest	&	OpenI	&	PadChest	
    &	OpenI	&	PadChest	&	OpenI	&	PadChest	
    &	OpenI	&	PadChest     &	OpenI	&	PadChest  
    \\	\midrule
        Atelectasis & 86.04 & 85.23 & 83.80 & 85.46 & 85.34 & 84.74 & 87.27 & 85.18 & 88.23 & 83.00 & 87.91 & 86.19 \\ 
        Cardiomegaly & 91.42 & 92.12 & 89.42 & 91.45 & 88.08 & 89.17 & 84.59 & 89.83 & 89.12 & 91.40 & 91.37 & 92.17 \\ 
        Effusion & 95.58 & 96.19 & 93.96 & 95.89 & 94.54 & 95.63 & 94.28 & 96.75 & 93.67 & 96.36 & 95.28 & 96.71 \\ 
        Infiltration & 68.76 & 64.08 & 60.48 & 67.98 & 72.26 & 74.20 & 76.10 & 76.19 & 82.08 & 71.27 & 81.65 & 76.64 \\ 
        Mass & 80.20 & 86.04 & 85.00 & 85.98 & 88.08 & 80.56 & 82.79 & 84.80 & 75.12 & 80.67 & 92.31 & 88.48 \\ 
        Nodule & 87.60 & 75.68 & 85.12 & 75.60 & 86.44 & 74.82 & 83.42 & 75.27 & 82.10 & 74.34 & 84.05 & 75.28 \\ 
        Pneumonia & 91.01 & 76.87 & 88.87 & 76.40 & 83.50 & 76.17 & 87.18 & 78.20 & 85.65 & 74.83 & 89.99 & 78.71 \\ 
        Pneumothorax & 84.28 & 79.22 & 83.07 & 76.98 & 74.07 & 76.10 & 75.89 & 80.02 & 73.93 & 76.45 & 88.89 & 85.82 \\ 
        Edema & 82.27 & 92.40 & 85.66 & 93.87 & 83.38 & 88.23 & 87.31 & 89.55 & 85.92 & 93.01 & 87.60 & 98.68 \\ 
        Emphysema & 82.05 & 80.87 & 82.36 & 75.80 & 76.94 & 73.10 & 80.94 & 78.15 & 75.16 & 74.21 & 85.28 & 81.94 \\ 
        Fibrosis & 87.53 & 91.50 & 90.67 & 94.57 & 92.09 & 96.43 & 90.19 & 95.35 & 91.06 & 95.29 & 94.56 & 97.44 \\ 
        Pleural Thicken & 87.37 & 76.06 & 82.66 & 76.62 & 82.83 & 72.82 & 84.12 & 70.55 & 80.10 & 68.14 & 86.94 & 71.53 \\ 
        Hernia & 96.60 & 94.17 & 94.69 & 92.74 & 80.85 & 70.11 & 91.95 & 85.84 & 87.29 & 81.38 & 98.57 & 94.22 \\ 
        Mean AUC & 86.21 & 83.88 & 85.06 & 83.79 & 83.72 & 80.93 & 85.08 & 83.51 & 83.80 & 81.56 & 89.57 & 86.45 \\ 
        \bottomrule
    \end{tabular}
    }
    \vspace{-10pt}
\end{table*}

\section{Additional Results}
\subsection{Per-finding results}
We show per-finding results over all available findings for NIH~\cite{nih} in Tables~\ref{tab:nih_result_1} and~\ref{tab:nih_result_2} and for CheXpert~\cite{cxp} in Tables~\ref{tab:cxp_result_1} and~\ref{tab:cxp_result_2} .

\subsection{Hyper-parameter sensitivity}
\cref{tab:ablation} studies the four hyper-parameters ($\lambda$, $\gamma$, $M$ and $K$) of BoMD.
In general, for $\lambda$, we note that relying too much on the pseudo-labels from the graph ($\lambda=0.2$) or the original noisy labels ($\lambda=1.0$) worsens the performance, with the best result achieved with a balanced $\lambda=0.6$.
We noticed that the method is robust to $\gamma$ and $M$ with little variation in results.
As for $K$, values larger than 10 over-smooth the decision boundary of our classifier, causing under-fitting. 
The values $\lambda=0.6$ and $\gamma=0.25$, $M=3$, and $K=10$ reach the best results. 

\subsection{Evaluation for Descriptors from MID}
\begin{figure*}
    \centering
    \includegraphics[width=.33\linewidth]{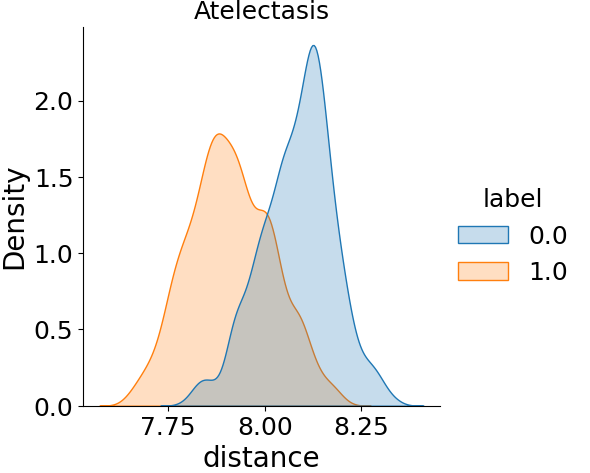}
    \includegraphics[width=.33\linewidth]{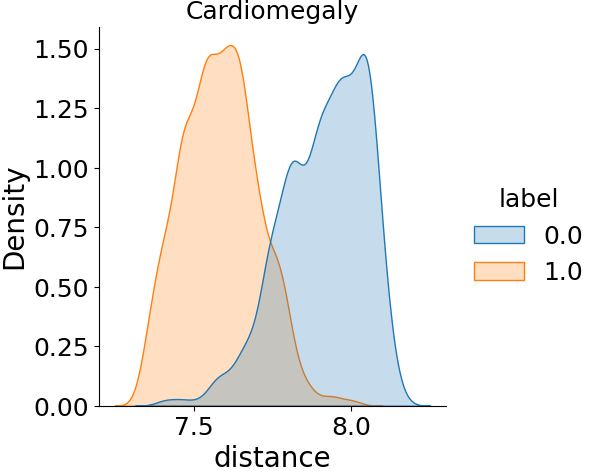}
    \includegraphics[width=.33\linewidth]{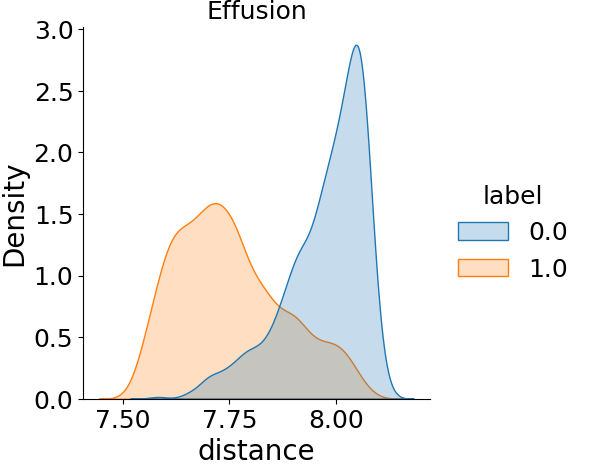}
    \\
    \includegraphics[width=.33\linewidth]{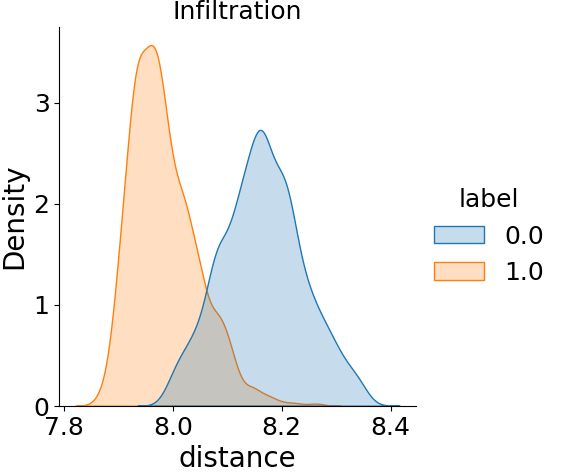}
    \includegraphics[width=.33\linewidth]{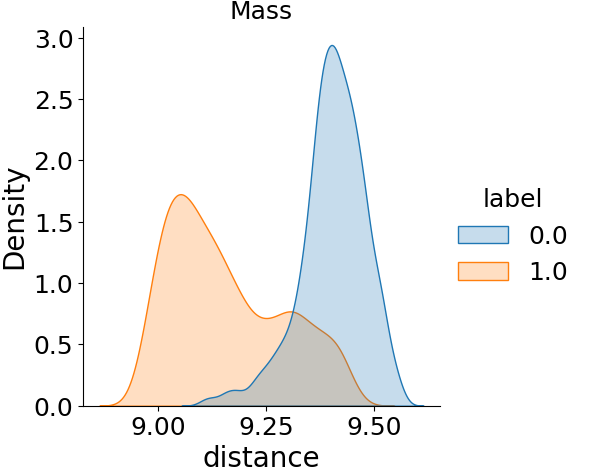}
    \includegraphics[width=.33\linewidth]{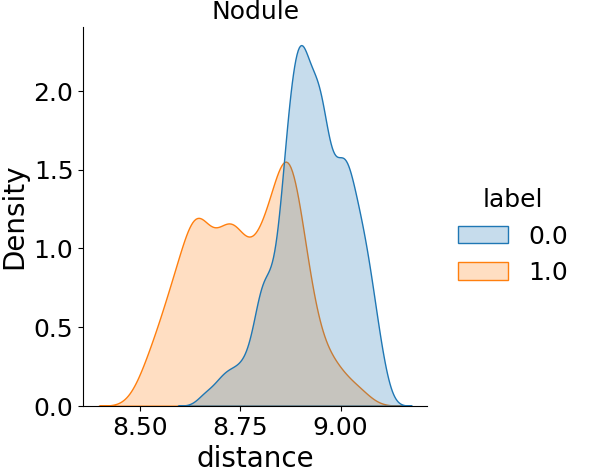}
    \\
    \includegraphics[width=.33\linewidth]{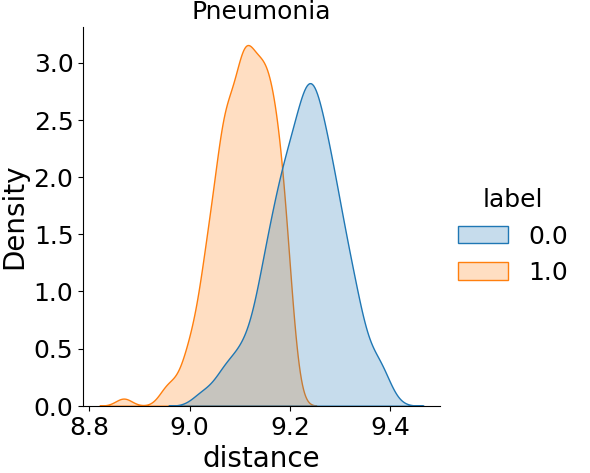}
    \includegraphics[width=.33\linewidth]{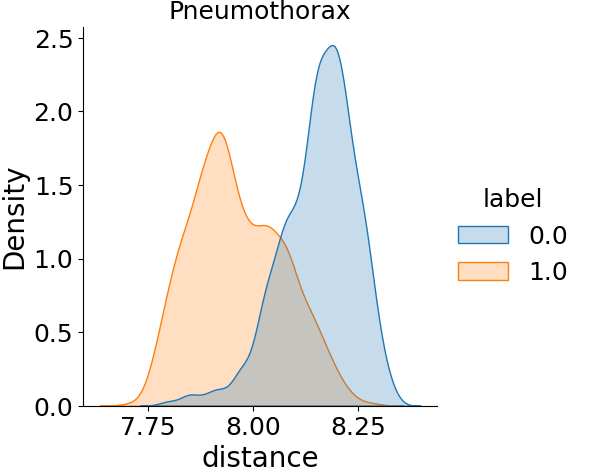}
    \includegraphics[width=.33\linewidth]{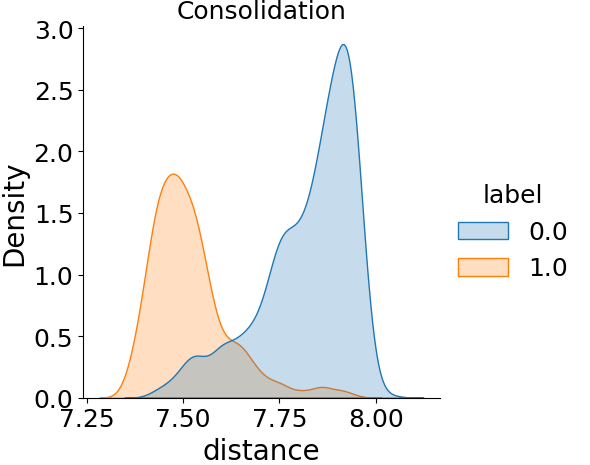}
    \\
    \includegraphics[width=.33\linewidth]{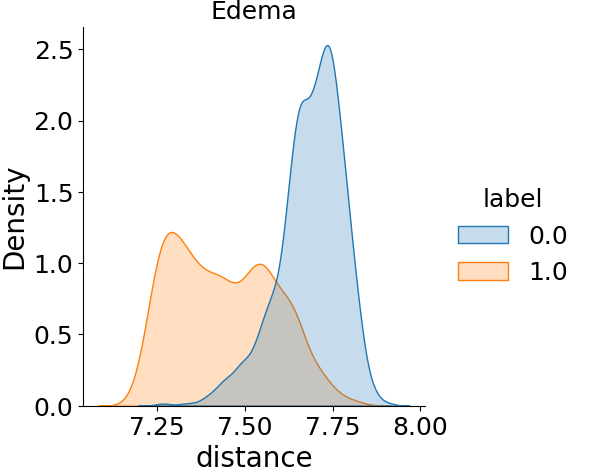}
    \includegraphics[width=.33\linewidth]{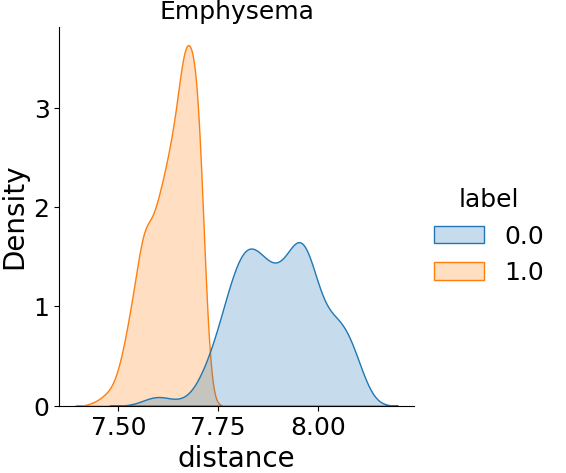}
    \includegraphics[width=.33\linewidth]{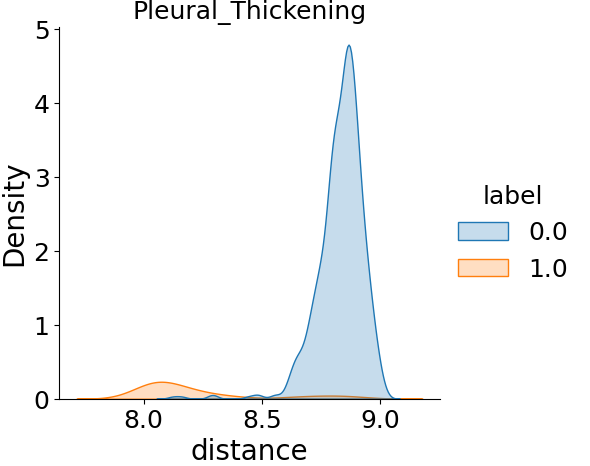}
    \caption{L2 distance between positive/negative descriptors and semantic descriptor}
    \label{fig:descriptor_score}
\end{figure*}

\noindent\textbf{Visualisation of distance distribution.}
To verify the separation of positive descriptors (labelled as 1) and negative descriptors (labelled as 0) based on their edge weight, we performed an analysis on a dataset consisting of 12 classes. Each class contained 4,000 samples, along with its corresponding semantic descriptors from the NIH dataset~\cite{nih}.
For each class, we denote positive samples' descriptors as ``1'', and negative samples' descriptors as ``0''.
The analysis involved examining the distribution of L2 distance, and the results are presented in Figure~\ref{fig:descriptor_score}. Our findings suggest that, on average, positive descriptors are 
closer to their corresponding semantic descriptors than negative descriptors, which proves the effectiveness of our MID module.

\begin{figure*}
    \centering
    \includegraphics[width=.33\linewidth]{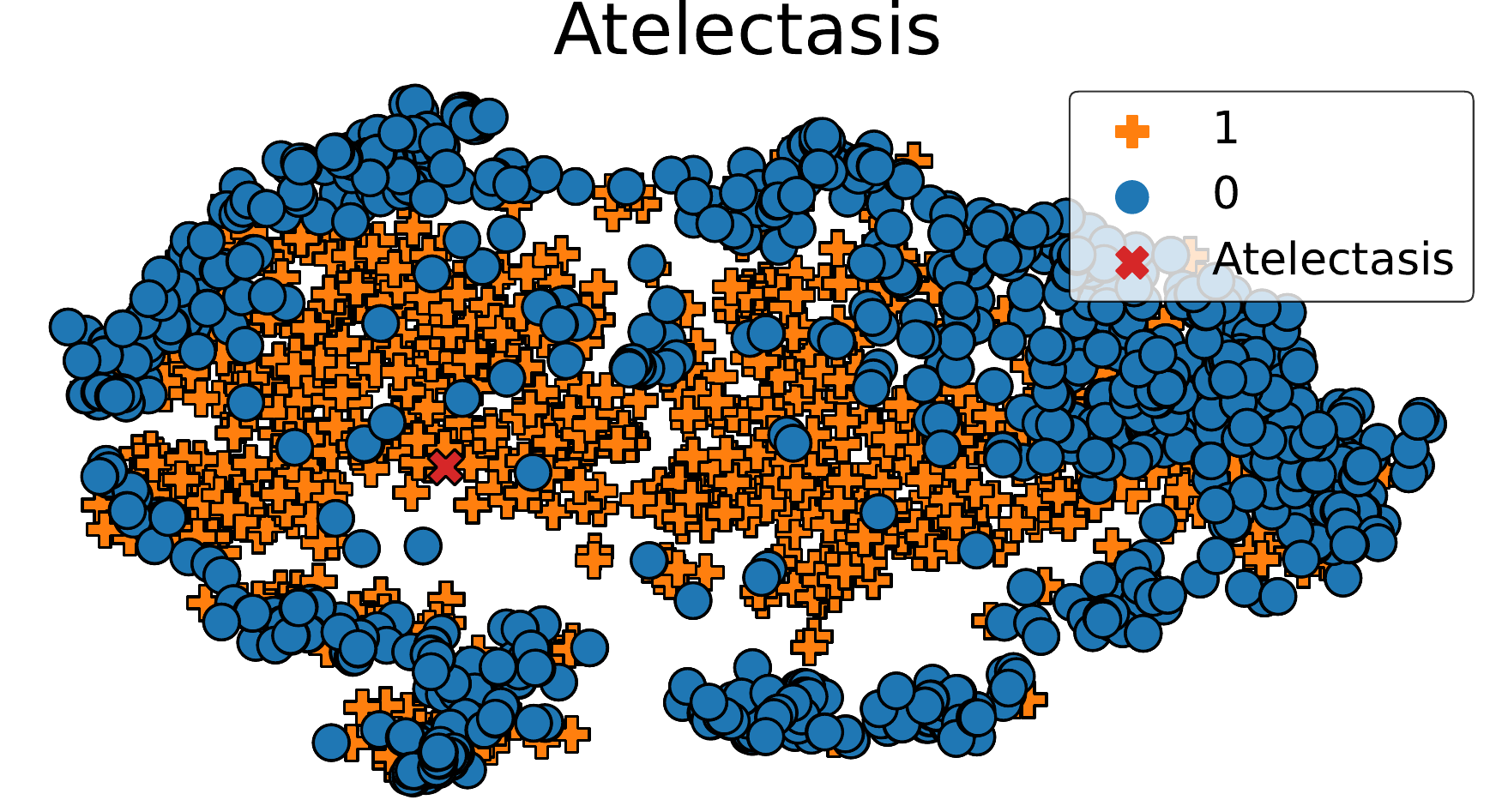}
    \includegraphics[width=.33\linewidth]{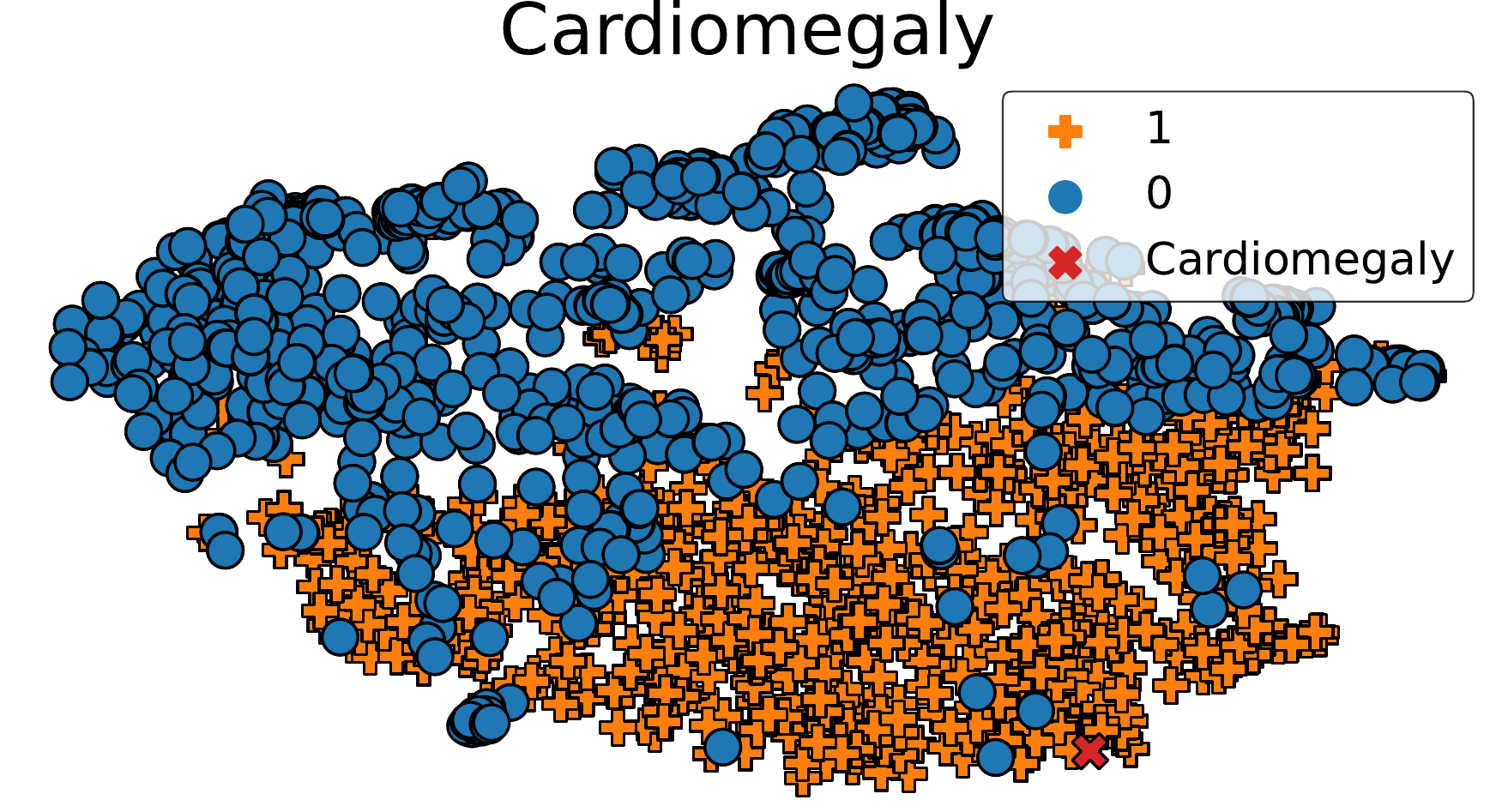}
    \includegraphics[width=.33\linewidth]{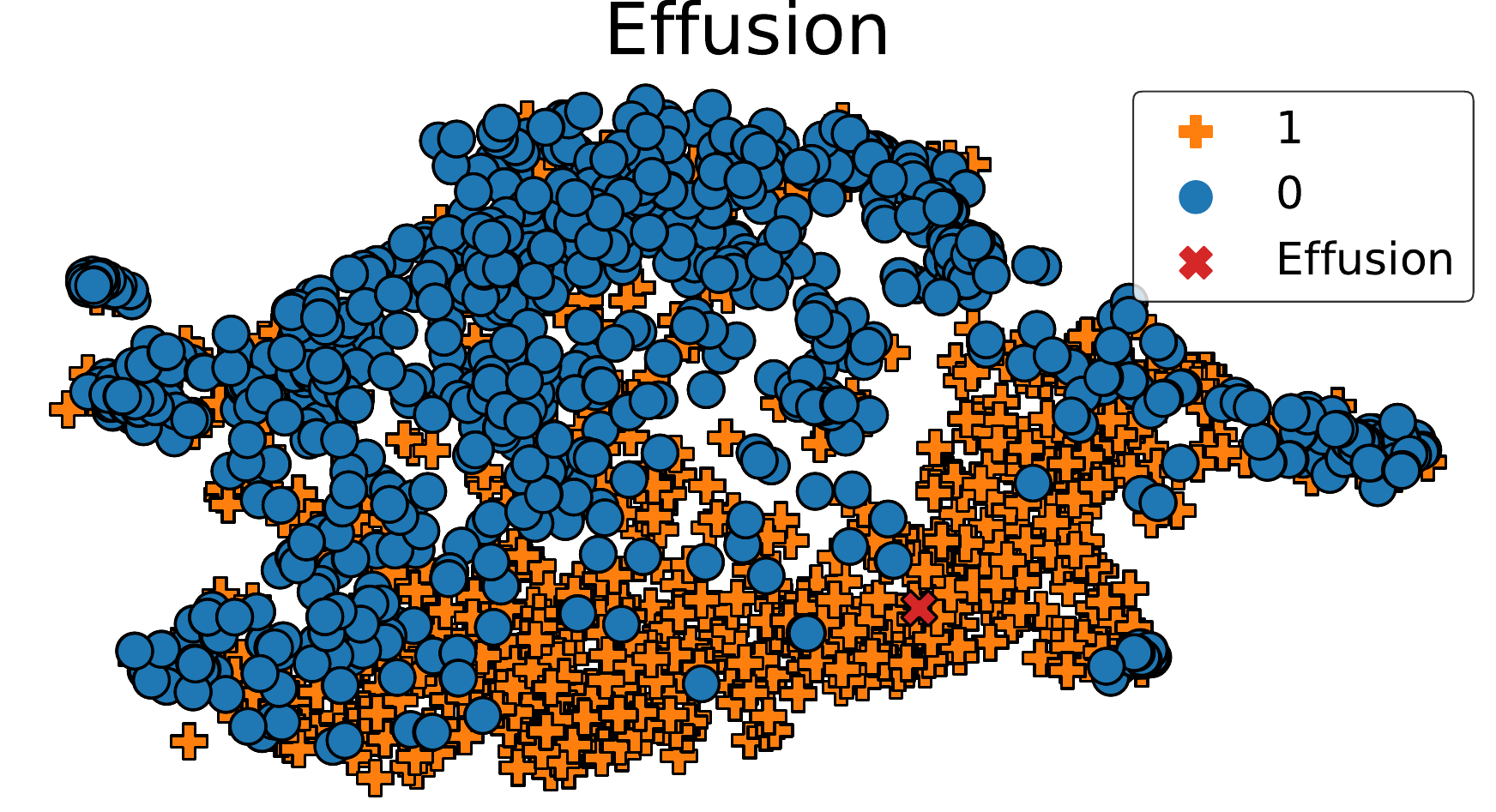}
    \vspace{10pt} \\
    \includegraphics[width=.33\linewidth]{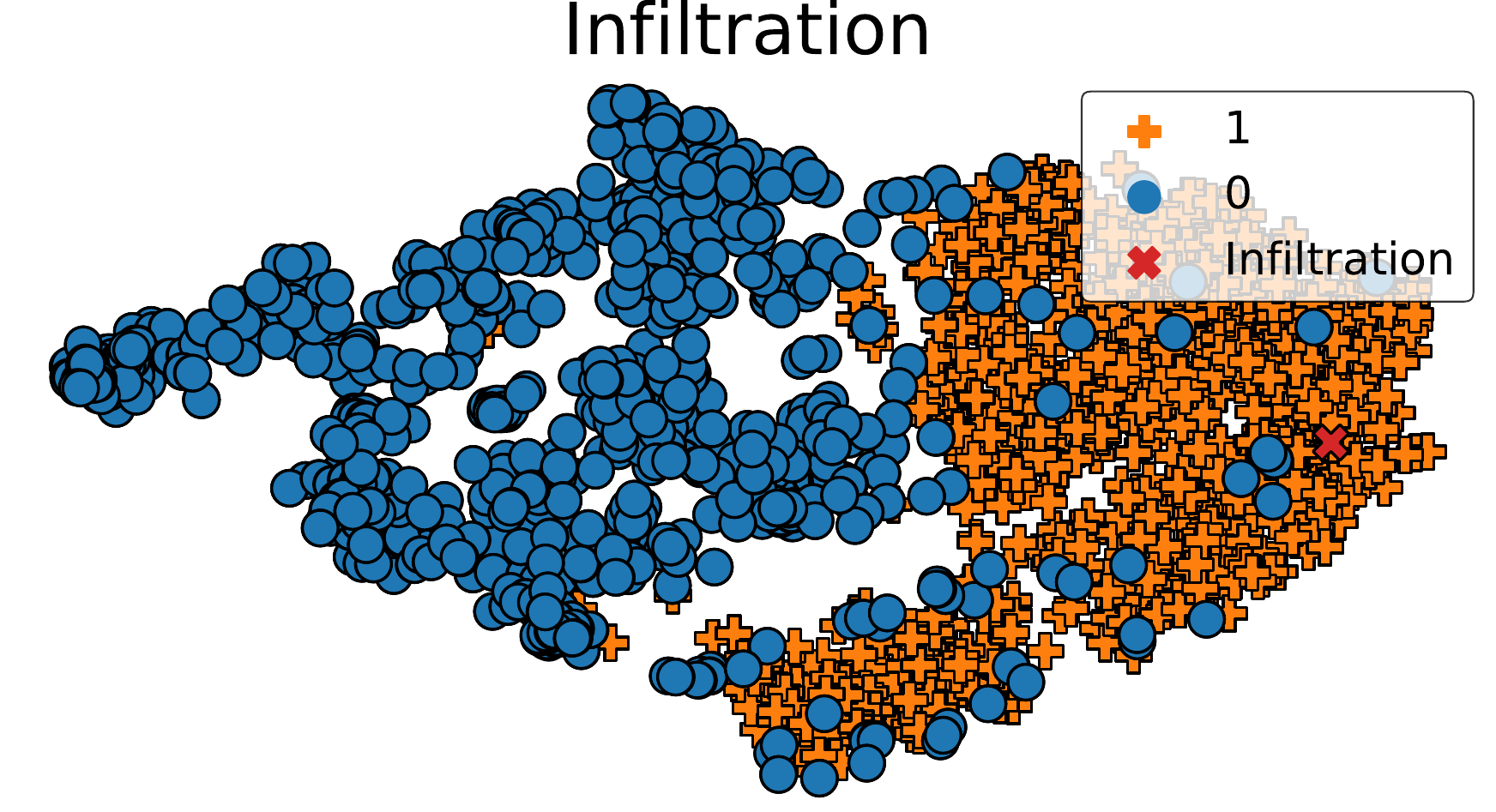}
    \includegraphics[width=.33\linewidth]{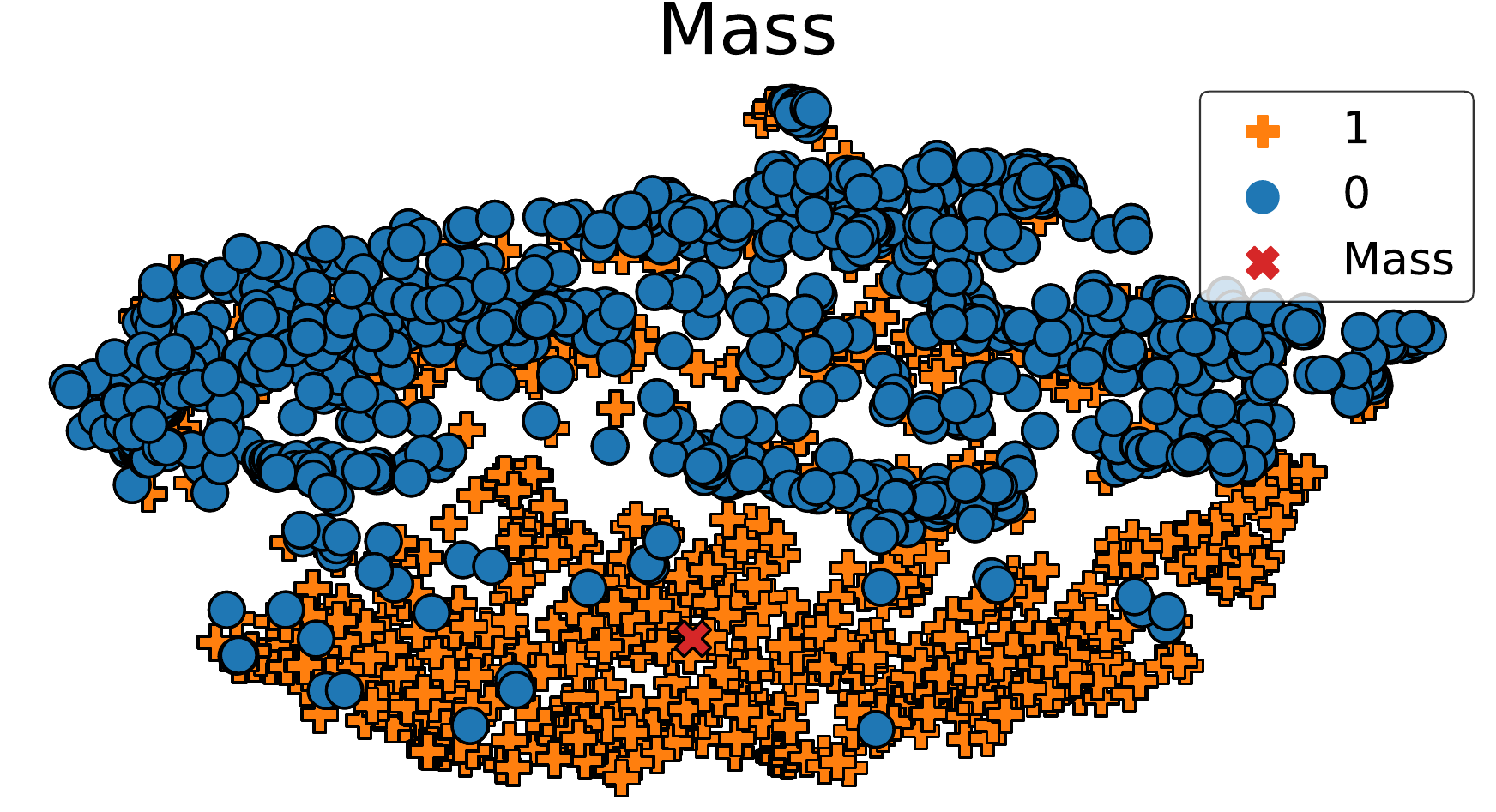}
    \includegraphics[width=.33\linewidth]{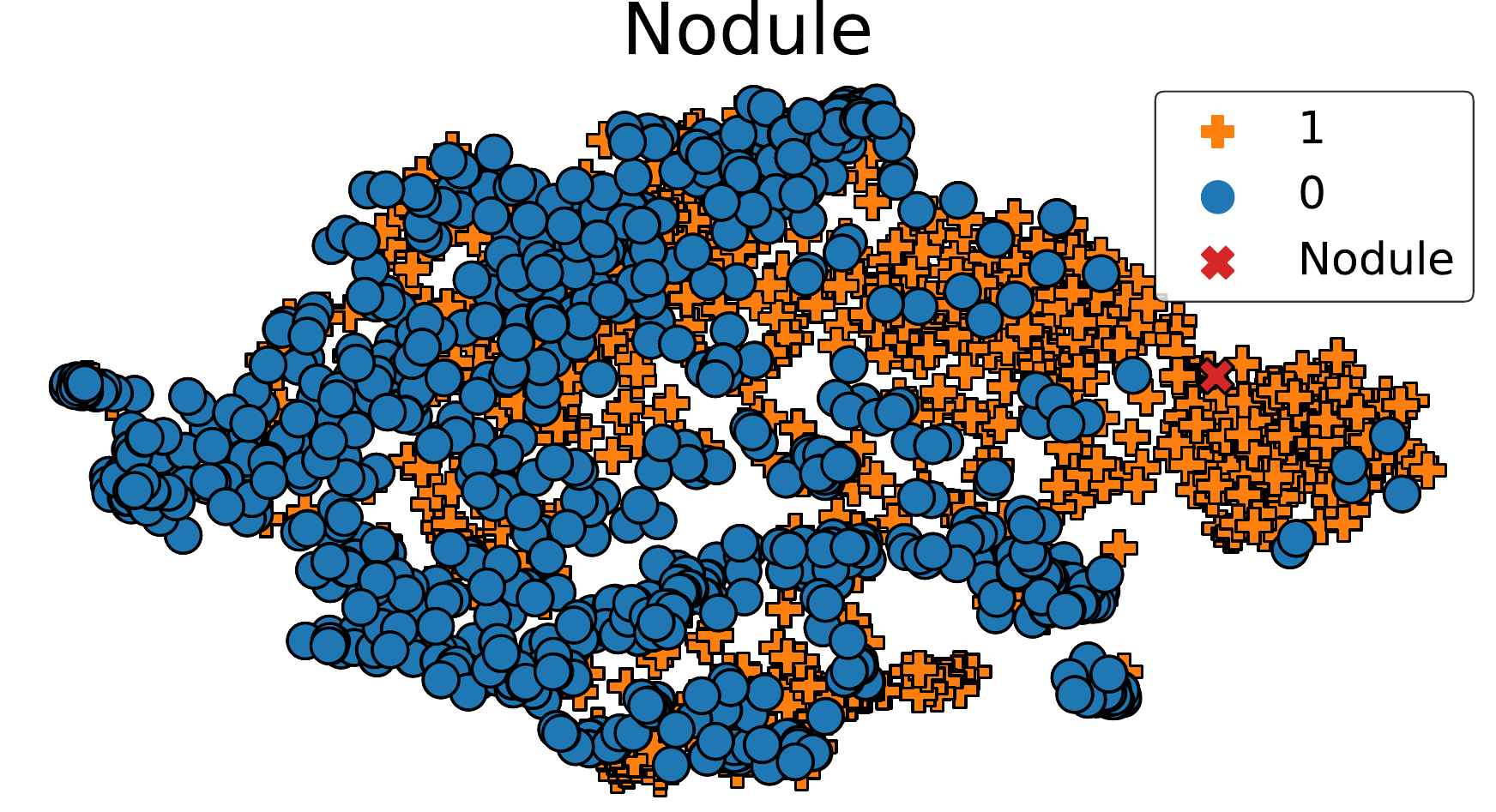}
    \vspace{10pt} \\
    \includegraphics[width=.33\linewidth]{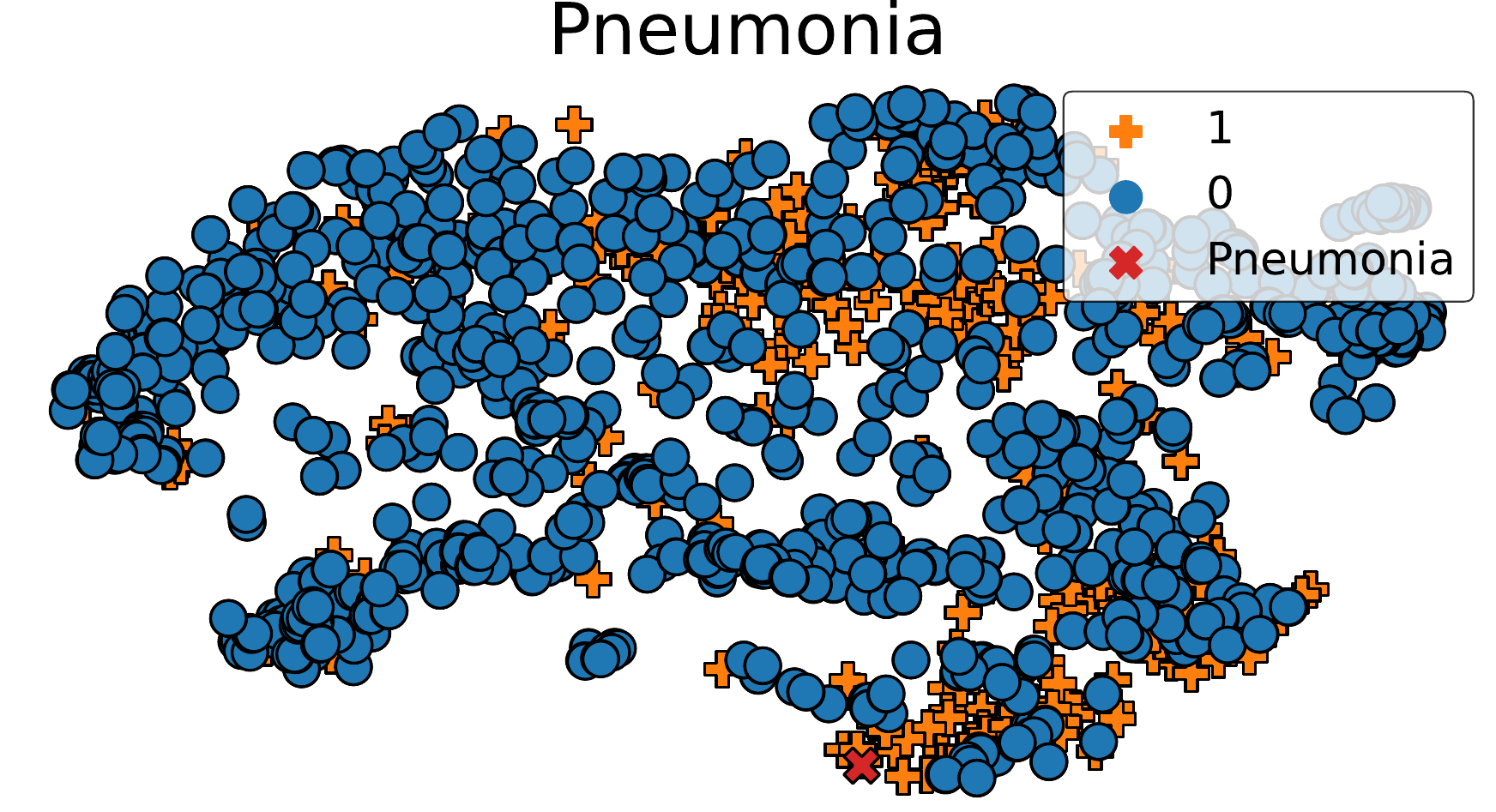}
    \includegraphics[width=.33\linewidth]{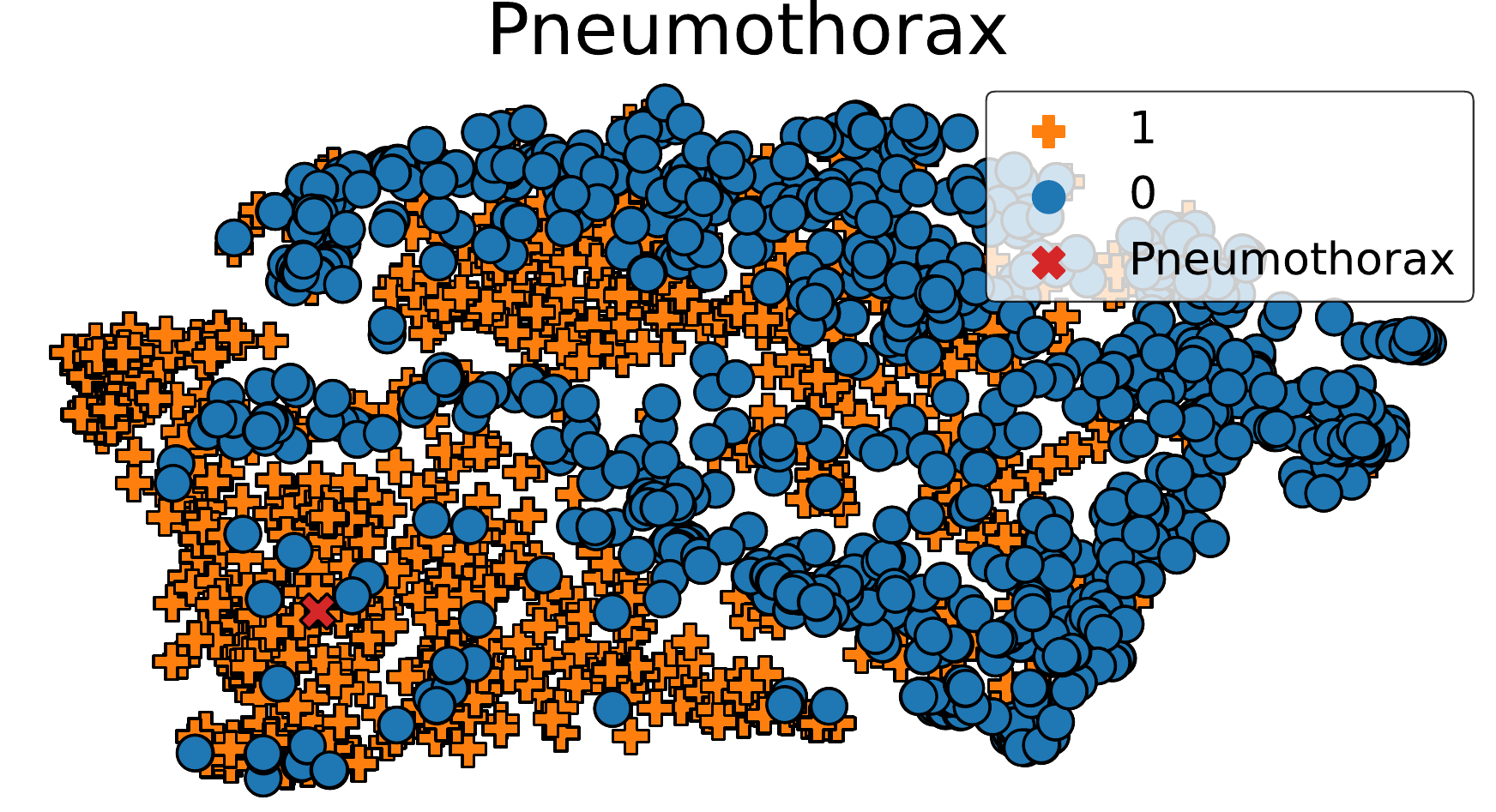}
    \includegraphics[width=.33\linewidth]{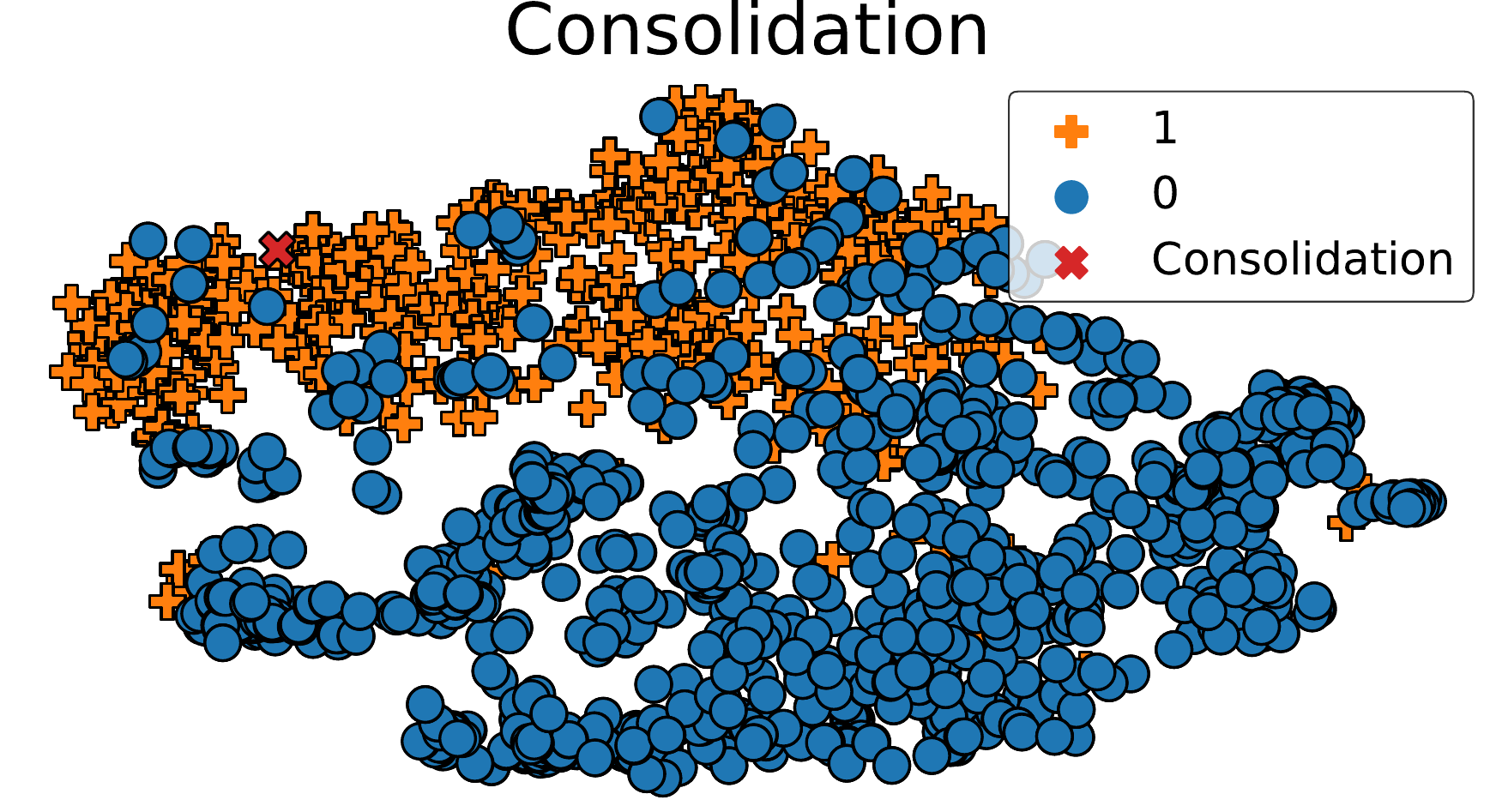}
    \vspace{10pt} \\
    \includegraphics[width=.33\linewidth]{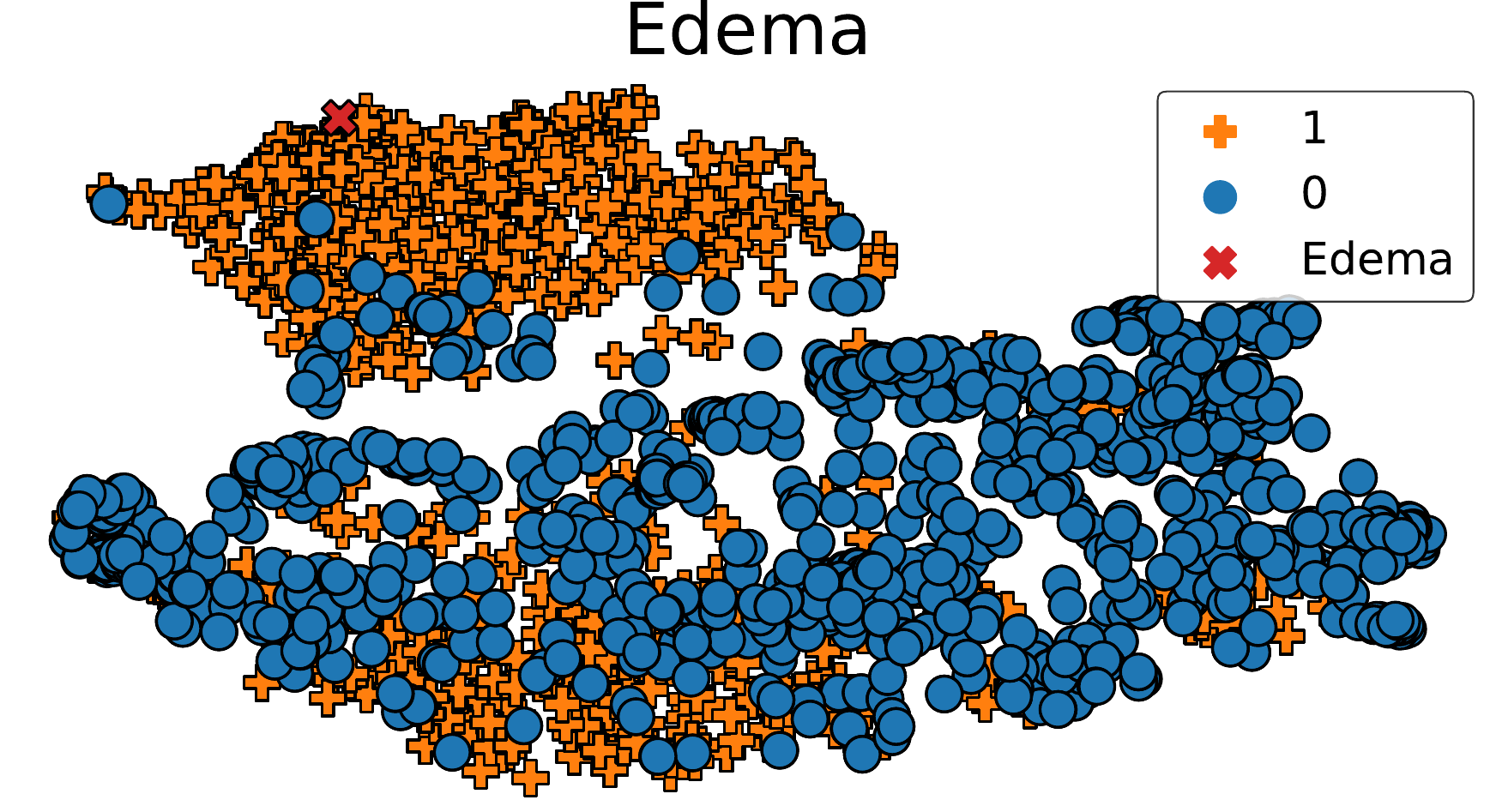}
    \includegraphics[width=.33\linewidth]{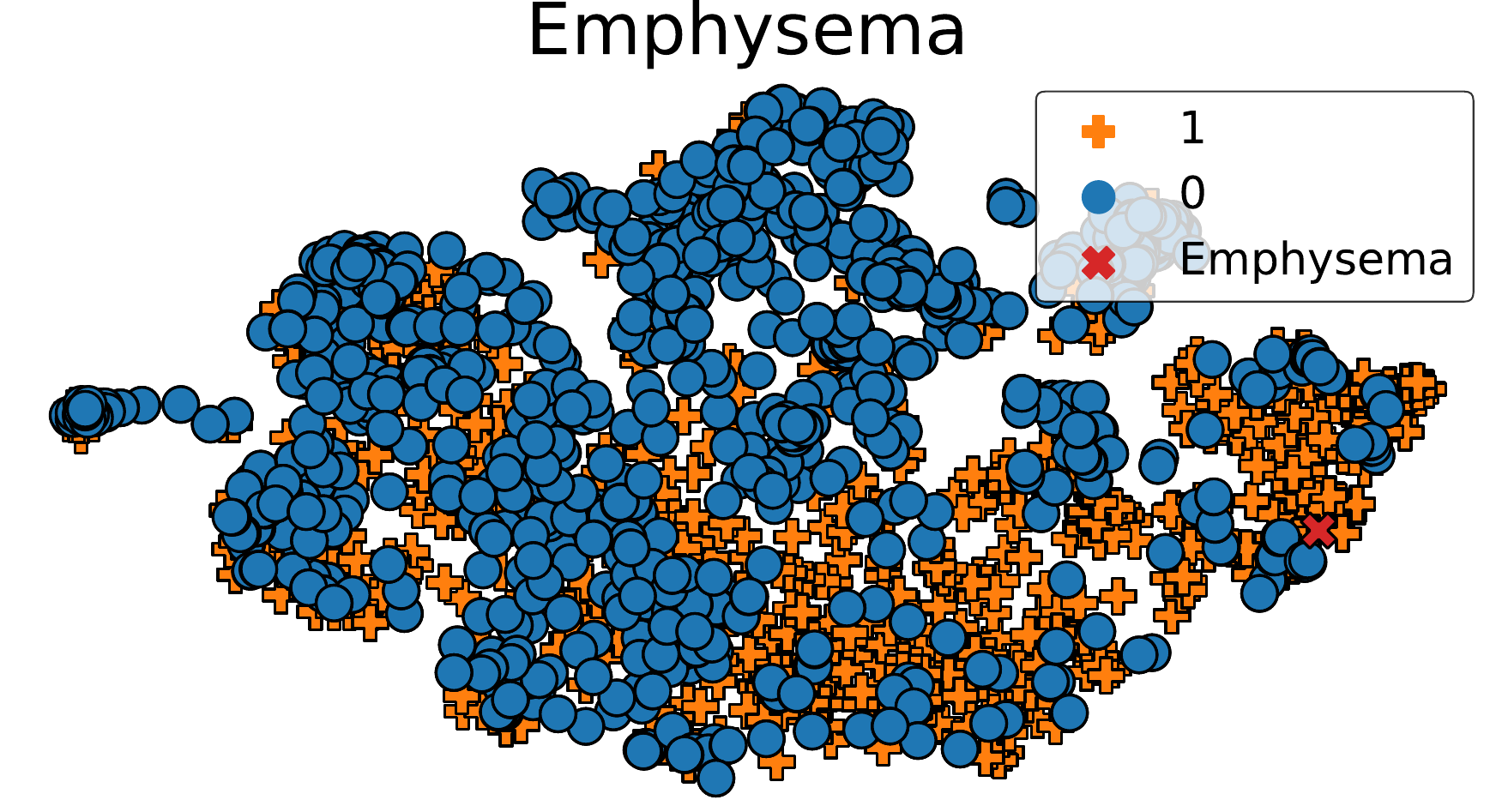}
    \includegraphics[width=.33\linewidth]{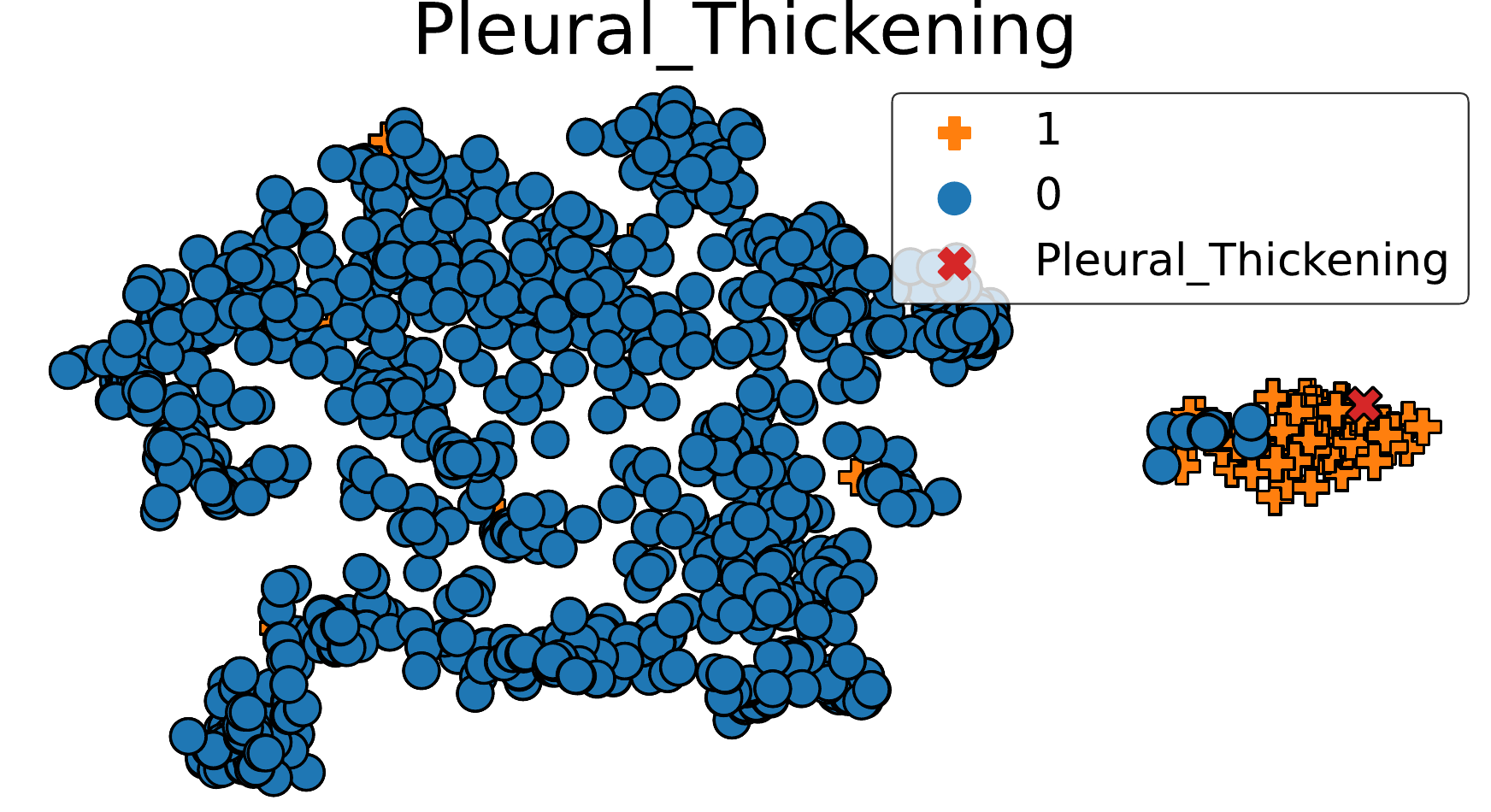}
    \caption{Visualisation of descriptor distribution in latent space.}
    \label{fig:descriptor_tsne}
\end{figure*}

\noindent\textbf{Visualisation of latent space.}
To visualise the descriptors' distribution in the latent space, we plot the t-SNE~\cite{van2008visualizing} for 12 classes with 4,000 samples per class sampled from NIH~\cite{nih}, as shown in Fig.~\ref{fig:descriptor_tsne}. For each class, we denote positive samples' descriptors as $+$, negative samples' descriptors as $\circ$ and semantic descriptors as $\times$. We show that the semantic descriptors are mostly surrounded by class-related descriptors ($+$), which varied the clustering effect of our MID module. Such clustering effect will benefit our graph-based smooth re-labelling as shown in Sec~\ref{sec:visual_ls}

\begin{table*}[ht]
    \centering
    \caption{Disease-level testing AUC results for models that \textbf{trained on CheXpert}. 
    }
    \vspace{-5pt}
    \label{tab:cxp_result_1}
    \resizebox{1\textwidth}{!}{  
    \begin{tabular}{l|cc|cc|cc|cc|cc|cc}
    \toprule
    Models	
    &	\multicolumn{2}{c|}{Hermoza et al}	&	\multicolumn{2}{c|}{CAN}	&	\multicolumn{2}{c|}{DivideMix}	    &	\multicolumn{2}{c|}{FINE}	
    &	\multicolumn{2}{c|}{ELR}	        &	\multicolumn{2}{c}{NVUM}	
    \\	\midrule	\midrule			
    Datasets	
    &	OpenI	&	PadChest	&	OpenI	&	PadChest	
    &	OpenI	&	PadChest	&	OpenI	&	PadChest	
    &	OpenI	&	PadChest     &	OpenI	&	PadChest
    \\	\midrule
        Cardiomegaly & 86.12 & 87.20 & 82.83 & 85.89 & 79.53 & 85.42 & 83.62 & 83.99 & 90.48 & 87.46 & 85.15 & 88.48 \\
        Edema & 87.92 & 94.35 & 86.46 & 97.47 & 81.24 & 83.41 & 86.43 & 87.07 & 90.88 & 96.12 & 87.35 & 97.21 \\ 
        Pneumonia & 65.56 & 57.15 & 61.88 & 63.38 & 55.98 & 51.20 & 55.58 & 55.58 & 61.59 & 64.13 & 64.42 & 67.89 \\ 
        Atelectasis & 78.40 & 75.65 & 80.13 & 72.87 & 72.74 & 68.34 & 72.87 & 72.87 & 79.63 & 73.68 & 80.81 & 75.03 \\ 
        Pneumothorax & 62.09 & 78.65 & 74.69 & 79.50 & 75.49 & 79.98 & 65.34 & 68.85 & 74.12 & 83.95 & 82.18 & 83.32 \\ 
        Effusion & 87.00 & 93.94 & 88.43 & 92.92 & 83.75 & 88.91 & 85.92 & 85.92 & 86.65 & 92.42 & 83.54 & 89.74 \\ 
        Fracture & 57.47 & 53.77 & 59.96 & 60.44 & 63.87 & 62.23 & 51.97 & 62.50 & 56.75 & 62.00 & 57.02 & 62.67 \\ 
        Mean AUC & 74.94 & 77.24 & 76.34 & 78.92 & 73.23 & 74.21 & 71.68 & 73.83 & 77.16 & 79.97 & 77.21 & 80.62 \\ 
        \bottomrule									
    \end{tabular}
    }
    \vspace{-10pt}
\end{table*}

\begin{table*}[ht]
    \centering
    \caption{Disease-level testing AUC results for models that \textbf{trained on CheXpert}. 
    }
    \vspace{-5pt}
    \label{tab:cxp_result_2}
    \resizebox{1\textwidth}{!}{  
    \begin{tabular}{l|cc|cc|cc|cc|cc|cc}
    \toprule
    Models	
    &	\multicolumn{2}{c|}{NPC}	        &	\multicolumn{2}{c|}{NCR}	
    &	\multicolumn{2}{c|}{LS}	            &	\multicolumn{2}{c|}{OLS}	
    &	\multicolumn{2}{c|}{GLS}	        &	\multicolumn{2}{c}{\textbf{BoMD}}
    \\	\midrule	\midrule			
    Datasets	
    &	OpenI	&	PadChest	&	OpenI	&	PadChest	
    &	OpenI	&	PadChest	&	OpenI	&	PadChest	
    &	OpenI	&	PadChest     &	OpenI	&	PadChest
    \\	\midrule
        Cardiomegaly & 80.33 & 86.43 & 90.10 & 86.84 & 85.53 & 83.42 & 83.58 & 86.29 & 88.22 & 87.30 & 90.85 & 89.88 \\ 
        Edema & 82.35 & 79.09 & 90.11 & 98.26 & 89.72 & 99.43 & 85.17 & 95.69 & 87.92 & 97.49 & 89.89 & 98.76 \\ 
        Pneumonia & 62.31 & 64.52 & 58.80 & 59.87 & 49.64 & 50.41 & 64.18 & 56.48 & 59.49 & 63.64 & 65.35 & 66.10 \\ 
        Atelectasis & 81.29 & 76.13 & 79.01 & 72.22 & 75.13 & 69.30 & 70.85 & 71.75 & 76.71 & 73.32 & 80.01 & 74.33 \\ 
        Pneumothorax & 82.32 & 82.35 & 78.06 & 86.15 & 73.05 & 78.33 & 80.10 & 83.36 & 77.53 & 77.58 & 82.99 & 86.04 \\ 
        Effusion & 78.71 & 86.65 & 85.62 & 91.57 & 84.70 & 90.97 & 84.64 & 91.83 & 85.19 & 91.94 & 87.37 & 93.07 \\ 
        Fracture & 59.92 & 65.95 & 56.80 & 60.63 & 52.27 & 55.52 & 67.13 & 58.60 & 60.44 & 60.32 & 63.72 & 64.12 \\ 
        Mean AUC & 75.32 & 77.30 & 76.93 & 79.36 & 72.86 & 75.34 & 76.52 & 77.72 & 76.50 & 78.80 & 80.03 & 81.76 \\ 
        \bottomrule
    \end{tabular}
    }
    \vspace{-10pt}
\end{table*}

\end{document}